**Evaluation of the uncertainty in calculating nanodosimetric quantities due to the use of different interaction cross sections in Monte Carlo track structure codes.**

*Carmen Villagrasa[1*], Giorgio Baiocco[2], Zine-El-Abidine Chaoui[3], Michael Dingfelder [4], Sébastien Incerti[5,] Pavel Kundrát[6], Ioanna Kyriakou[7], Yusuke Matsuya[8,9], Takeshi Kai[9], Alessio Paris[10], Yann Perrot[1], Marcin Pietrzak[11,12], Jan Schuemann[13] and Hans Rabus[14].*

[1]Autorité de sûreté nucléaire et de radioprotection (ASNR), PSE-SANTE/SDOS/LDRI, F-92260, Fontenay aux Roses, France

[2] Radiation Biophysics and Radiobiology Group, Physics Department, University of Pavia, Pavia, Italy

[3]Physics Department, Faculty of Sciences, University Setif1. 19000. Algeria

[4] Department of Physics, East Carolina University, Greenville, North Carolina, USA

[5]Université de Bordeaux, CNRS, LP2I, UMR 5797, F-33170 Gradignan, France

[6]Department of Radiation Dosimetry, Nuclear Physics Institute of the CAS, Na Truhlářce 39/64, 180 00 Praha, Czech Republic

[7]Medical Physics Laboratory, Department of Medicine, University of Ioannina, Ioannina 45110, Greece

[8]Faculty of Health Sciences, Hokkaido University, Kita-12 Nishi-5, Kita-ku, Sapporo, Hokkaido 060-0812, Japan

[9]Nuclear Science and Engineering Center, Japan Atomic Energy Agency, 2-4 Shirane Shirakata, Tokai, Ibaraki 319-1195, Japan

[10]Department of Radiation Oncology, Mayo Clinic, Jacksonville, Florida, USA

[11] Institut Curie - Proton Therapy Center of Orsay, Radiation Oncology Department, 15 Rue Georges Clémenceau, 91400 Orsay, France

[12]Institut Curie, PSL Research University, University Paris Saclay, INSERM LITO, Campus universitaire, 91898 Orsay, France

[13] Department of Radiation Oncology, Massachusetts General Hospital & Harvard Medical School, Boston, MA 02114, USA

[14]Physikalisch-Technische Bundesanstalt (PTB), Abbestrasse 2-12, 10587 Berlin, Germany

*Corresponding author

E-Mail: carmen.villagrasa@asnr.fr

## Abstract

This study evaluates the uncertainty in nanodosimetric calculations caused by variations in interaction cross sections within Monte Carlo Track Structure (MCTS) simulation codes. Nanodosimetry, essential for understanding the biological effects after exposure to ionizing radiation, relies on accurately simulating particle interactions at the molecular scale. Different MCTS codes, developed independently over decades, employ distinct physical models and datasets for electron interactions in liquid water, a surrogate for biological tissues. The paper focuses on the Ionization Cluster Size Distribution (ICSD) generated by electrons of varying energies in nanometric volumes. Seven MCTS codes (several options in Geant4-DNA, PARTRAC, PHITS-TS, MCwater and PTra) were tested using their native cross sections and a common dataset derived from averaging data used in the participating codes. The results reveal significant discrepancies among the codes in ICSDs and derived biologically relevant nanodosimetric quantities such as mean ionization numbers ($M_1$) and probabilities of obtaining two or more ionizations ($F_2$). The largest variations were observed for low-energy electrons, where the contribution from interaction cross sections dominates the overall uncertainties. For instance, $M_1$ values for ICSDs of electron of 20 eV can differ by around 45 % (RSD) and 34 % (RSD) was found for $F_2$ values of ICSDs of electrons of 50 eV. Using common cross sections substantially reduced the discrepancies, suggesting that cross section datasets are the primary source of variability. Finally, estimates of deoxyribonucleic acid (DNA) damage using the PARTRAC code highlight tht cross section variations have a non-negligible impact simulated biological outcomes, particularly for double-strand breaks (DSBs) Indeed, despite the fact that many other parameters in the simulation that can greatly differ from one code to another, the different interaction cross-sections studied in this work can lead to differences in the number of DSBs calculated with the PARTRAC code of up to 15%.

## Introduction

Ionizing radiation elicits both beneficial and detrimental effects in biological systems. It is widely used in medicine for both diagnostic (e.g., X-ray, CT, or PET imaging) and therapeutic purposes (e.g., radiotherapy for the treatment of cancer), in industry (e.g., defectoscopy), and in science. Avoiding adverse radiation effects as far as reasonably achievable is the general aim of radiation protection. Biological effects of diverse types of ionizing radiation differ widely. Accounting for these differences is crucial in numerous fields, including ion radiotherapy (Durante et al., 2021; Scholz, 2020), nuclear medicine (Li et al. 2018), or radiation protection in space (ICRP, 2013; Cuccinotta et al. 2003; Edwards 2001; Schulte et al., 2008; Zaider 1996). Mechanistic models and simulations have largely succeeded in explaining these differences and tracking them to the underlying mechanisms of radiation physics, chemistry, and biology.

This mechanistic approach starts by describing as precisely as possible the processes that occur at the molecular level within exposed cells. Diverse models then proceed towards deriving the effects on larger scales, such as damage to cellular deoxyribonucleic acid (DNA) or even later biological endpoints such as mutation induction or cell inactivation (Nikjoo et al. 1994, Krämer and Scholz 2006; Friedland et al. 2011a, 2011b, Meylan et al. 2017, Matsuya et al. 2019; Scholz et al. 2020; Shin et al. 2021, Chatzipapas et al. 2024, Le Tuhan et al. 2024). In general, this mechanistic approach uses numerical simulation methods, the first step of which is to obtain a nanometric-scale description of the physical interactions of ionizing radiation with the biological material. The energy imparted in these interactions results in molecular damage, particularly damage to nuclear DNA. The damage pattern is governed by the pattern of energy transfer points (including those from the primary particle and all its secondaries), known as a particle track.

Various simulation codes are available for this type of simulation, known as Monte Carlo Track Structure (MCTS) codes, reviewed e.g. in (Kyriakou et al. 2022). In Monte Carlo codes for dosimetric applications, particle transport in matter is simulated using the condensed-history approach. In this approach, charged-particle transport is based on multiple scattering processes and continuous slowing down approximations to save computing time. In contrast, MCTS codes perform a discrete step-by-step simulation of all interactions of the primary particle and all secondary (and higher-generation) electrons down to very low kinetic energies, typically a few electron volts (eV). The probabilities of interactions occurring are obtained from the corresponding interaction cross sections.

Different MCTS codes have been developed over the last four to five decades by various research groups; some but not all of them are publicly available (Kyriakou et al. 2022). Depending on the main objective of their development, these codes differ in the types of radiation they can transport and in the target materials used. Most of these codes can simulate electron tracks in liquid water, which is often used as a surrogate for biological material and has the properties of a condensed medium. The latter is indeed an important condition for these codes to be used in radiobiology since, at low energies, the interaction probabilities depend strongly on the phase (condensation state) of the target material. Most codes can simulate further particle types, frequently including photons, protons, and heavier ions. Some codes have been developed dedicatedly to accompany experimental nanodosimetry and detector design (Grosswendt et al. 2002; Bug et al. 2013). These codes can simulate tracks of electrons, protons or alpha particles in tissue-equivalent gases such as nitrogen or propane. Other codes use theoretical models to calculate the interaction cross sections of electrons or other charged particles in DNA bases (Bug et al. 2017, Swadia et al. 2017; Zein et al. 2023).

Many of the codes were developed in the last decades of the 20$^{th}$ century as a part of research carried out by individual groups or laboratories and, therefore, are not always publicly available. However, several general Monte Carlo codes that are publicly available have extended their scope over the past fifteen years by adding track-structure capabilities, often limited to electron transport in liquid water. As a result, their large user communities can benefit from these new options for a variety of applications, so that this type of calculation is being democratized in the scientific community. In particular, this is the case for the code series Geant4 (GEometry ANd Tracking), PHITS (Particle and Heavy Ion Transport code System), and MCNP (Monte Carlo N Particle). Geant4 (Agostinelli et al. 2003, Allison et al. 2006, 2016) has its Geant4-DNA extension (Bernal et al. 2015, Incerti et al. 2010a, 2010b, 2018, Tran et al. 2024), also available in derived codes such as TOPAS/TOPAS n-Bio (Faddegon et al. 2020) and GATE (Sarrut et al. 2022). Similarly, PHITS has been extended by track-structure modes (PHITS-TS) (Matsuya et al. 2019,2022, Sato et al. 2024). A module is also available in MCNP for the same purpose; however, low-energy cross section data have been incorporated into MCNP only for cold atomic targets and may differ significantly due to not considering molecular and other low-energy physics (Kulesza et al. 2022).

While most MCTS codes can simulate electron tracks in liquid water, the cross sections employed may differ between codes. This is because different theoretical models were derived from the few experimental data available (Heller et al. 1974; Hayashi et al. 2000). Therefore, results obtained for different simulated quantities characterizing track structure in microdosimetry and nanodosimetry can differ greatly depending on which MCTS has been used (Villagrasa et al. 2022).

All the underlying cross section models have been validated as far as possible by comparing the simulation results with more integrated data, such as stopping powers or particle ranges [ICRU 1984; 2014]. However, measurement of lineal energy or ionization clusters in microscopic or nanoscopic volumes directly in water is impossible with present technology, so benchmarking the corresponding simulation results against comparably detailed experimental data is not feasible. The use of gas detectors to this aim, implies the use of water cross-sections for its material-density scaling procedure (Grosswendt et al. 2004), and the experimental uncertainties finally obtained are high (Pietrzak et al. 2021). Therefore, assessing the degree of systematic uncertainty of the results obtained with any particular code is nowadays difficult.

Working group 6 (WG 6) of the European Radiation Dosimetry Group (EURADOS e.V.) is concerned with quality assurance of computational methods used in radiation dosimetry (Rabus et al. 2022). Within WG6, task 6.2 is dedicated to micro- and nano- dosimetry. The present work is part of an activity of task 6.2 to study the dispersion of results for different nanodosimetric quantities obtained with diverse codes. The goal is to better understand the origin of these discrepancies, especially to disentangle contributions due to different cross section datasets used in various codes for modelling low-energy electron transport. The focus on low-energy electrons is motivated by their unique role in energy deposition and biological effects produced by diverse radiation types, such as high-energy electrons, photons, protons, alpha particles, or heavier ions. For instance, secondary electrons with energies below 5 keV account for about a third of the energy imparted by $^{60}$Co gamma-rays or a half for 220 kVp X-rays. These low-energy electrons are mainly responsible for the induction of DNA double-strand breaks (NCRP, 2018).

This paper presents results from analysing the simulated frequency distributions of the number of ionizations produced by low energy electrons in a nanometric volume (ionization cluster size distribution, ICSD). This quantity has been chosen in this work for two main reasons: mathematically, it is very sensitive to differences in the track structure codes used, and from a physical perspective, ICSD is a fundamental quantity of nanodosimetry that can also be experimentally measured, at least in gas-based detectors (Bantsar at al. 2018), and that is often used to characterize different radiation qualities in terms of their biological impact. Indeed, from ICSDs it has been possible to estimate the yields of DNA double-strand breaks and other types of DNA damage (Conte et al. 2023). Therefore, to quantify the dispersion of simulated ICSDs and their derived nanodosimetric quantities due to the different assumptions on interaction cross sections for low-energy electrons, the different MCTS codes were used to simulate tracks of 20 eV – 10 keV electrons in liquid water. Subsequently, the code-specific cross sections were replaced with a common set of cross sections and used by each code to re-calculate the ICSDs. The results enable assessing the uncertainty in nanodosimetric quantities due to the use of alternative MCTS and the underlying cross sections. Finally, to illustrate the potential impact of the differences in cross section data used by diverse MCTS codes on observables closer to biological effects than ICSDs, DNA damage simulations have been performed with one of the codes, PARTRAC, using both its original and the common cross section datasets.

## Materials and Methods

### Monte Carlo Track structure codes participating

Gathering a representative set of MCTS codes is critical to be able to quantify how the variability of underpinning cross section values impacts predictions of different MCTS codes. EURADOS Task Group 6.2 involves various researchers with access to some of the MCTS codes used in the field. With a great deal of publicity and information sharing about this activity, further researchers became interested in the topic and willing to dedicate their time to make progress on this issue. This activity has compiled a set of results that may be considered sufficiently representative of the MCTS codes actively used in nanodosimetry and related areas. The following codes were included in the exercise: Geant4-DNA (with models in so-called option 2, option 4 and option 6) (Bernal et al. 2015, Incerti et al. 2010a, 2010b, 2018, Tran et al. 2024), PHITS-TS (Matsuya et al. 2022, Sato et al. 2024), PARTRAC (Friedland et al. 2011a), PTra (Physikalisch-Technische Bundesanstalt track structure code) (Grosswendt 2002; Grosswendt et al. 2004; Bug et al. 2013) and MCwater, an inhouse MCTS code from the Sétif University, that uses published models to account for both inelastic and elastic interactions of electrons in liquid water (Penn et al. 1976, 1987, Shinotsuka et al. 2017, Salvat et al. 2005).

The inelastic and elastic interaction cross section models used by each of these codes for electron transport in liquid water are summarized in Table 1 and the corresponding references.

Table 1: MCTS codes used in this work and references to the physical models used to calculate interaction probabilities of electrons in liquid water within these codes

| Code | Ionization | Excitation | Elastic scattering |
| --- | --- | --- | --- |
| Geant4-DNA-Opt 2 | PWBA [1] (Drude functions) [a,*] | PWBA [1] (same as ionizations) | Partial-wave formalism [2] |
| Geant4-DNA-Opt 4 | PWBA [1] (Drude function algorithm) [b,*] | PWBA via the Kramers-Kronig relation [1] (same as ionization) | Uehara screened Rutherford model [3] |
| Geant4-DNA-Opt 6 | Relativistic BEB (CPA-100) [4] | PWBA [5] (same as PARTRAC)[**] | Independent Atom Method (IAM) (from the CPA100 code) [4] for T> 50 eV and experimental data on ice for T<50 eV [6] |
| PARTRAC | PWBA [5] (Drude function with Kramers-Kronig relation) [c,*] | PWBA [5] (same as ionization)[**] | NIST ELAST database values for atomic hydrogen and oxygen [7] |
| MCwater | PWBA [8,9] (full Penn algorithm) [d] | PWBA [8,9] (same as ionization) | ELSEPA code [10] with electronic densities of [11] |
| PTra | BEB [12] + K shell from [13] | Discrete values from [14] | T <200 eV: [15] T> 200 eV: [16] |
| PHITS | T < 100 keV: Discrete values [17]; T >100 keV : Relativistic BEB [12] | T < 100 keV : Discrete values [17] ; T >100 keV : PWBA [18] | Moliere's Model [19] |

PWBA: plane-wave Born approximation, BEB: Binary encounter Bethe model, CPA100: MC code of Terrisol et al. (1990)
[1] (Heller et al. 1974 ; Kyriakou et al. 2015),[2] Champion et al. 2009, [3] Uehara et al.1992,[4] Bordage et al. 2016, [5] (Dingfelder et al. 1998, 2008), [6] Michaud and Sanche, 1987 , [7] Berger et al.1993,

[8] (Dingfelder et al. 2014, Hayashi et al. 2000), [9] Penn, D. R.,1987,[10] Salvat et al. 2005, [11] Neuefeind et al. (2002), [12] Kim et al. 2000, [13] Kaplan et al. (1990),[14] Bigildeev and Michalik, 1996,[15] Brenner and Zaider,1983, [16] Grosswendt and Waibel,1978,[17] Kai et al., 2024,[18] Paretzke 1988, [19] Moliere 1948

a Optical data model based on Drude-like functions.
b Optical data model based on Drude-like functions with a different parametrization algorithm. In the present work an extended relativistic version of this algorithm has been used described in Kyriakou et al., 2025.
c Optical data model based on Drude-like functions with a different representation of the optical data.
d Optical data model based on the Penn model.
* For K-shell ionization different models are used as described in the corresponding references.
**Different Drude parametrizations are used that can give rise to quite different cross section values between the codes (Kyriakou et al. 2015)

It can be noted that the Geant4-DNA code appears three times with different options. Indeed, more than a code, Geant4 and its extension Geant4-DNA are MC toolkits that enable users to build their own track structure simulation. In Geant4-DNA, different models are available to simulate the transport of low-energy electrons, regarding both inelastic and elastic interactions. From these different models, the developers have recommended alternative "constructors" or options, which, for this work, means that 3 independent calculations are available in terms of interaction cross sections.

It is important to highlight that the aim of this article is not to compare the models used to calculate the interaction cross sections in the different codes or to judge them as more or less accurate. The information on these models is given here to help identify MCTSs that use similar, or very different models, in order to better understand the origin of the dispersion of results that will be shown in the next section.

## Simulated configuration and quantities

The simulation configuration used to assess the contribution of the different interaction cross sections to the dispersion of ICSDs and their derived nanodosimetric quantities was kept simple, to prevent participants' misunderstanding of the setup which could lead to additional dispersion in the results. For comparability, the simulation geometry was chosen like the one used in an earlier study (Villagrasa et al. 2019). This configuration is close to simulation scenarios that are often used in this type of calculation, so that the results obtained could be directly applied for instance with sparsely ionizing radiation where electron spurs are well separated or with radionuclides emitting low-energy electrons. It must be noted, however, that the interaction producing the initial electron is not considered, or in the case of radionuclides, the cation produced when electrons are emitted is not followed in this work.

The participants were asked to calculate the ICSDs produced in a nanometric sphere of liquid water by monoenergetic electrons starting at its centre. The target sphere was immersed in liquid water. Its diameter was 8 nm for electron energies of 20, 50, 100, 300, 600 and 1000 eV, and 100 nm for 5 and 10 keV. These volumes were chosen, as explained above, close to the ones used in previous studies and with diameters corresponding to the same order of magnitude of biological targets: nucleosome / DSB cluster dimensions for the low-energy electrons, and chromatin loop dimensions for the highest energies studied in this work. All codes were used in their track-structure mode using the lowest possible electron energy cutoff (down to which the electron interactions are explicitly simulated), which was between 1 and 10 eV, depending on the code. The number of events ($10^5$) was recommended to have a statistical uncertainty well below 5%.

From the results on the ICSDs in the 8 or 100-nm diameter spheres sent by the participants, different nanodosimetric quantities were calculated and compared, namely, the first momentum of the ICSD distribution, i.e., the mean number of ionizations ($M_1$), and the cumulative probability of obtaining 2 or more ionizations ($F_2$) and three or more ionizations ($F_3$), i.e. the probabilities of obtaining an ionization cluster or a complex ionization cluster (Conte et al. 2018).

## Modification of the MCTS codes

The methodology used in this exercise was straightforward but required the participants to modify their respective MCTS code. Therefore, only researchers having access to the source code and a deep knowledge of how their code works could perform this activity. Indeed, participants were first asked to extract interaction cross section values conventionally used by their code for interactions of electrons in liquid water, namely impact ionizations, electronic excitations and elastic scattering. A template was circulated among the participants to fill in these cross sections, indicating: the electron kinetic energy, the cross sections values for ionization of the 5 water molecular orbitals and for the 5 electronic excitations of the water molecule typically considered in these codes, and the total elastic interaction cross section.

For some codes/models, this procedure was facilitated by the fact that the codes use interaction cross section data tables. In that case, the task reduced to use the interpolation methods in the code to extract from the implemented tables the values corresponding to those requested in the template. In other cases, however, when the codes calculate the cross sections directly from mathematical formulae, these formulae had to be extracted from the source code to enable the calculation of the values requested for the table.

After compiling the templates filled by the initial participants in the work (some other joined afterwards), mean values for the different interaction cross sections were calculated for each electron energy and each ionization and excitation channel. The resulting cross sections are referred to as "common cross sections".

Because these common cross sections were calculated before the MCwater code and the current public version (ver. 3.34) of PHITS-TS joined the project, those were not included in their calculation. Nevertheless, a former version of the PHITS-TS code cross section was considered in the common set. Therefore, the common-set values for inelastic scattering of electrons of a given kinetic energy were calculated from the mean values in Geant4-DNA options 2, 4 and 6, PTra, PARTRAC and the relativistic BEB values included in this former version of PHITS. For elastic scattering, the common cross sections were determined from the models included in the same codes and values from the ELSEPA model (also available in Geant4-DNA) as additional data. It should be noted that the method used in this work to evaluate the variation in the dispersion of results requires codes to be modified and use common cross sections, but it is not necessary for these common cross sections to be the exact average of the cross sections of all participating codes. Therefore, it was not deemed necessary to re-evaluate these common cross sections when two new codes or code versions were added to the exercise.

The common cross sections were then distributed to the participants, who were asked to implement them in their codes (in a tabular form), run the ICSD simulations again, and report the new results.

The initial value of the total ionization, excitation and elastic cross sections used by the participating MCTS and the corresponding common-set of cross sections are shown in Supplementary Figures S1-S3.

Important remarks concerning this procedure that must be considered when analysing the results are:

-In the case of inelastic cross sections (ionization and electronic excitation), the angular distribution of the electrons ('projectile' or 'secondary' for ionizations) after the interaction was kept as in the original MCTS code. Only the interaction probability was modified.

-The same applies to elastic scattering: only the interaction probability for a given electron energy was prescribed by the common cross section dataset, not the angle of the electron after the elastic collision (differential elastic cross sections were kept as included in the given MCTS codes originally, but their influence has been studied, see Supplementary Figure S4).

- Single differential cross sections for ionization interactions were not modified. The probability of a given kinetic energy for the secondary electron (and thus for the projectile, as there is energy and momentum conservation) after the ionization of a given shell was kept as in the original codes.

-The specific interpolation methods/algorithms used in the different MCTS to calculate the interaction cross sections from the data tables were kept unmodified

Therefore, the residual dispersion in the new results obtained with all the MCTS codes applying the same common cross section data set will be due to all these factors. Its quantification will inform of their final importance in the nanodosimetric results obtained.

## Quantitative assessment of the differences between ICSDs

As mentioned in the introduction, ICSDs are often used to characterize radiation qualities in terms of the biological consequences they can produce, and, more specifically, in terms of DNA damage. To this end, various quantities are calculated from the ICSDs: the mean or first moment ($M_1$), the cumulative probability of having 2 or more ionizations in the volume ($F_2$), and of having 3 or more ionizations ($F_3$). The dispersion of these quantities has been assessed by the total and relative standard deviation of the sample of results obtained by the different participants with the original and the common cross section datasets.

In addition, we also propose in this work another metric, the Wasserstein $W_1$ distance (Givens and Scott 1984), for the comparison of the ICSDs. Indeed, in previous work (Villagrasa et al. 2019), the dispersion of microdosimetric and nanodosimetric distributions was specified by considering the distribution of the results from different participants for each energy bin or number of ionizations. In addition, the pairwise correlation of datasets obtained by different participants was assessed by Pearson's correlation coefficient. In this work, the desired deviation is between results from one code and an average over the results of all participants for each ionization cluster size. As the mean over all participants is correlated with each of the datasets from which it was constructed, this methodology could not be applied. For the same reason, using the $\chi^2$ metric is also not suited.

In this work, the Wasserstein-1 distances were used to quantify (as shown in Supplement 1):

a/ the "initial dispersion" of the MCTS codes by comparing each original ICSD to a mean ICSD calculated with all of them;

b/ the "final dispersion" of the modified MCTS codes by comparing the ICSD of each modified code to the mean ICSD calculated with all the modified codes;

c/ for each code the deviation of the ICSDs obtained with the original code and its modified version.

The initial dispersion between codes results from all the factors contributing to the calculation of ICSDs. On the other hand, the dispersion obtained after modifying the codes with the same interaction cross sections is due to all the other factors but that one. Thus, subtracting the two dispersions provides information on the influence of the modified cross sections on the ICSDs.

### Potential impact of cross section differences on predicted DNA damage

To illustrate how the different underlying cross section data may affect DNA damage yields and patterns predicted by diverse MCTS codes, corresponding simulations were performed with a single MCTS code, PARTRAC, using its original as well as the common cross section dataset.

These simulations used a spherical model of a lymphocyte cell nucleus with a diameter of 10 µm in which a multi-scale chromatin model was used. The model included an atomic description of the DNA, its winding around nucleosomes, formation of chromatin fibre, its loops, spherical chromatin domains and chromosomes in human interphase nuclei (Friedland et al. 2011b); the total DNA content was 6.6 Gbp. Monoenergetic 100 eV or 10 keV electrons were started isotropically at homogeneously distributed random points within the nucleus. A 2.11 µm liquid water shell around the nucleus was included to allow for potential re-entering of backscattered electrons into the nucleus, as used in previous simulations for ions with a tangential source disc (Friedland et al. 2017). Both direct and indirect effects were considered, i.e. damage induction through direct energy deposition to the DNA molecule and indirectly via the action of water radiolysis products, with standard assumptions and parameter values (Friedland et al. 2011a; 2017). For 100 eV and 10 keV electrons, respectively, in total $10^7$ and $10^6$ tracks were simulated in 100 runs (with $10^5$ or $10^4$ tracks per run); the particle numbers were chosen based on previous similar simulations to provide sufficient statistics within reasonable time (several hours CPU on a mid-class multi-thread laptop). The following classes of DNA damage were scored (Friedland et al. 2017): single-strand breaks (SSBs); double-strand breaks (DSBs, defined as strand breaks on opposite strands within 10 bp); DSB clusters (containing at least 2 DSB within 25 bp); and DSB sites (not discriminating between an isolated DSB and a DSB cluster).

## Results

### Diverse MCTS codes: dispersion in ICSDs

Figure 1 shows the ICSDs obtained by the participants for low-energy electrons in the 8 nm diameter liquid water spheres. Figure 2 displays results for the 100 nm diameter water sphere. Data can be found in Supplementary Table S- 1.

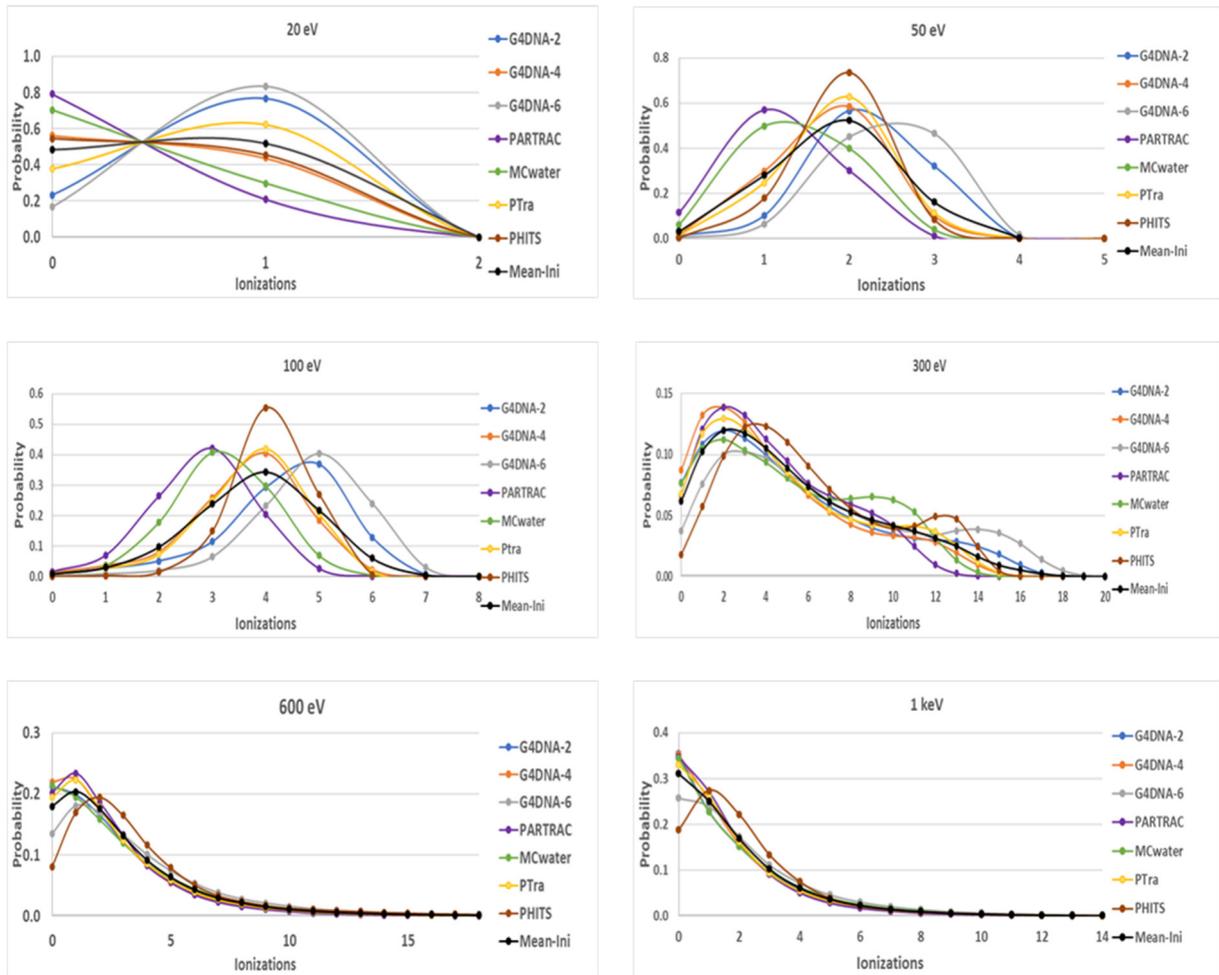

*Figure 1 : ICSDs obtained in a sphere of 8 nm in diameter made of liquid water and produced by a point source of mono-energetic electrons in its center. The initial electron energies are indicated in the graphs. The data were obtained by the diverse MCTS codes using their original interaction cross sections. The ICSD ("Mean-Ini") calculated as a mean frequency distribution of all the participants' ICSDs are shown in black. Note: the lines are only there to guide the eye, as non-integer values for the number of ionizations are meaningless.*

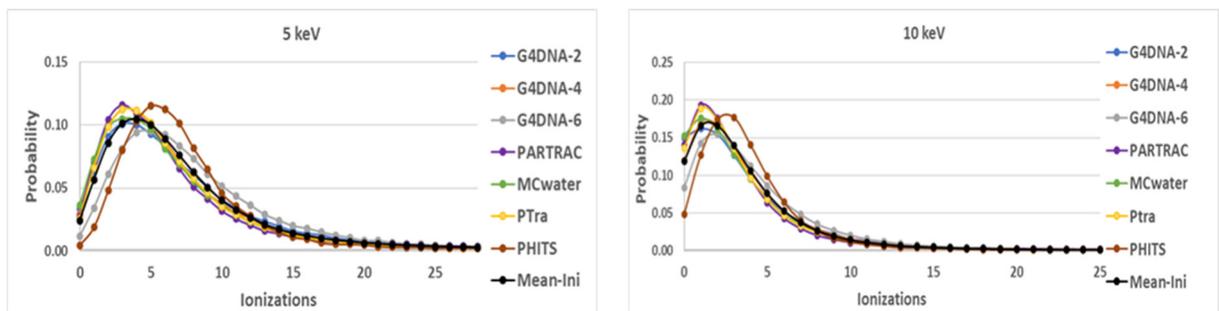

*Figure 2 : ICSDs obtained in a sphere of 100 nm in diameter made of liquid water containing, in its centre, a point source of mono-energetic electrons at the energies indicated in the graphs. The data were obtained by the diverse MCTS codes using their original interaction cross sections. The black symbols indicate the ICSD ("Mean-Ini") calculated as a mean frequency distribution of all the participants' ICSDs. Note: the lines are only there to guide the eye, as non-integer values for the number of ionizations are meaningless.*

The differences in these curves are evident, especially at low electron energies and for low ionization numbers in the volume. Comparison with the original cross section values of the codes shown in Supplementary Figs. S1 to S3 reveals a relationship between the ICSDs shown in

Fig. 1 and the differences between the cross sections used, especially for low energies (i.e., below 300 eV). For higher electron energies, a corresponding relationship is less evident. Hence the aim of this paper is to be able to quantify the contribution of the cross section values used for the different interactions in the codes in the observed differences. In Figs. 1 and 2, the black symbols represent the mean value at each ionization cluster size of all the ICSDs of the participating codes in their original version. This "mean" initial ICSD has been taken as a reference to calculate the Wasserstein-1 distances for each original frequency distribution. The mean value of all the calculated distances (see Supplementary Table S- 2 : Wasserstein distances calculated between each original ICSD and the mean ICSD presented in black color in Figure 1 and  Figure 2 Supplementary Table S- 1 Supplementary Table S- 2 : Wasserstein distances calculated between each original ICSD and the mean ICSD presented in black color in Figure 1 and  Figure 2) at a given electron energy is a measure of the initial dispersion among the codes. The corresponding values are given in  Table 2.

*Table 2 : Mean value of the Wasserstein-1 distances between the ICSDs from the original codes with respect to the Mean ICSD for each electron energy.*

| Energy (eV) | 20 | 50 | 100 | 300 | 600 | 1000 | 5000 | 10000 |
|---|---|---|---|---|---|---|---|---|
| **Mean value of the W-1 distances for original codes** | 0.19 | 0.35 | 0.56 | 0.66 | 0.40 | 0.22 | 0.70 | 0.40 |

Note that the values in Table 2 should not be overinterpreted in absolute terms and thus compared from one energy to another. Indeed, these values depend on the number of bins in the ICSDs which varies with energy, at least up to 600 eV, in our results (for the 8 nm diameter target).For energies of 5 and 10 keV, the volume in which the ICSDs have been calculated is different so that the absolute values of the average W-1 distances are not comparable with those at the smaller energies. Values in Table 2 are to be compared, energy per energy, with those obtained later in this paper with the modified codes (using the same calculation conditions for each energy).

## Dispersion in the nanodosimetric quantities from the original MCTS simulations

Table 3 and Table 4 give the values of the nanodosimetric quantities $M_1$ (mean number of ionizations, i.e., the mean value of the ICSD) and $F_2$ (probability of obtaining 2 or more ionizations in the target volume), respectively.

*Table 3 : Mean numbers of ionizations (first momenta of the ICSDs) in liquid water spheres of 8 nm diameter (for 20 eV – 1 keV electron energies) and 100 nm diameter (for 5 and 10 keV electron energies), respectively, obtained with the original MCTS codes. SD values refer to the standard deviations calculated from the results of the seven original MCTS codes (including the three alternative options in Geant4), RSD (relative standard deviation) values correspond to SD divided by the mean of these seven values.*

| $M_1$: Mean numbers of ionizations predicted by the original MCTS codes | | | | | | | | | |
|---|---|---|---|---|---|---|---|---|---|
| Energy (eV) | G4-DNA Opt 2 | G4-DNA Opt 4 | G4 -DNA Opt 6 | PARTRAC | MCwater | PTra | PHITS | SD | RSD |
| **20** | 0.77 | 0.44 | 0.83 | 0.21 | 0.30 | 0.62 | 0.45 | 0.23 | 0.45 |
| **50** | 2.21 | 1.77 | 2.43 | 1.21 | 1.42 | 1.84 | 1.90 | 0.42 | 0.23 |
| **100** | 4.30 | 3.65 | 4.83 | 2.81 | 3.17 | 3.72 | 4.09 | 0.68 | 0.18 |
| **300** | 5.35 | 4.55 | 6.65 | 4.44 | 5.21 | 5.01 | 6.04 | 0.79 | 0.15 |
| **600** | 2.90 | 2.54 | 3.59 | 2.50 | 2.99 | 2.77 | 3.76 | 0.49 | 0.16 |
| **1000** | 1.83 | 1.67 | 2.26 | 1.57 | 1.87 | 0.82 | 1.76 | 0.26 | 0.14 |
| **5000** | 7.10 | 6.56 | 8.09 | 6.20 | 6.28 | 6.38 | 6.86 | 0.66 | 0.10 |
| **10000** | 4.02 | 3.97 | 4.53 | 3.54 | 3.66 | 3.56 | 3.94 | 0.35 | 0.09 |

Table 4 : Values of the probability of obtaining more than 2 ionizations ($F_2$) in liquid water spheres of 8 nm diameter (for 20 eV-1 keV electron energies) and 100 nm diameter (for 5 and 10 keV electron energies), respectively, obtained from the ICSDs calculated with the original MCTS codes. SD values refer to standard deviations calculated from the results of the seven original MCTS codes (including the three alternative options in Geant4), RSD values correspond to SD divided by the mean of these seven values.

| $F_2$: Probability of two or more ionizations predicted by the original MCTS codes | | | | | | | | | |
|---|---|---|---|---|---|---|---|---|---|
| **Energy (eV)** | **G4-DNA Opt 2** | **G4-DNA Opt 4** | **G4-DNA Opt 6** | **PARTRAC** | **MCwater** | **PTra** | **PHITS** | **SD** | **RSD** |
| **50** | 0.89 | 0.69 | 0.93 | 0.31 | 0.74 | 0.44 | 0.82 | 0.23 | 0.34 |
| **100** | 0.96 | 0.95 | 0.99 | 0.92 | 0.96 | 0.95 | 1.00 | 0.03 | 0.03 |
| **300** | 0.81 | 0.78 | 0.89 | 0.81 | 0.82 | 0.82 | 0.92 | 0.05 | 0.06 |
| **600** | 0.59 | 0.56 | 0.68 | 0.57 | 0.58 | 0.59 | 0.75 | 0.07 | 0.12 |
| **1000** | 0.42 | 0.39 | 0.50 | 0.38 | 0.41 | 0.43 | 0.54 | 0.06 | 0.13 |
| **5000** | 0.90 | 0.90 | 0.95 | 0.86 | 0.86 | 0.89 | 0.92 | 0.033 | 0.04 |
| **10000** | 0.69 | 0.71 | 0.77 | 0.65 | 0.66 | 0.67 | 0.79 | 0.058 | 0.08 |

The relative standard deviation (RSD) allows us to compare values between different energies of the electrons. In the case of the mean number of ionizations, $M_1$, the dispersion is higher for low electron energies, being as high as 45% for the lowest one, 20 eV, where only 0 or 1 ionizations are possible. $M_1$ being the mean value of the ICSDs, the values in Table 3 can be understood from Figures 1 and 2. For instance, for 100 eV, ICSDs obtained with PARTRAC and Geant4-DNA opt6 are almost identical in shape but shifted by +2 ionizations. Hence, also $M1$ is shifted by 2 (2.81 and 4.83).

For the $F_2$ values (probability of two or more ionizations), the relative standard deviation is high for low energy (34% at 50 eV), but dramatically decreases at 100 eV where, even though the complete ICSDs are visibly different, the probabilities $P(0)$ and $P(1)$ of 0 and 1 ionizations are very close for all the codes. As $F_2$ is only affected by these probabilities (being $F_2 = 1-P(0)-P(1)$), its value is not representative of the full differences between ICSDs. The relative deviation of the $F_2$ values then increases with the electron energy for the constant volume of 8 nm diameter and decreases again when higher volumes are considered for 5 and 10 keV electrons, maintaining its relation with the relative contribution of $P(0)$ and $P(1)$ to the sum of the ICSDs (that are normalized to 1).

The values of $F_3$ (probability of three or more ionizations, given in Supplementary Table S- 5 : $F_3$ values obtained with original MCTS codes, and standard deviation (SD) and relative standard deviation (RSD) of this set of codes Supplementary Table S- 5 : $F_3$ values obtained with original MCTS codes, and standard deviation (SD) and relative standard deviation (RSD) of this set of codesSupplementary Table S- 5 : $F_3$ values obtained with original MCTS codes, and standard deviation (SD) and relative standard deviation (RSD) of this set of codesSupplementary Table S- 5 : $F_3$ values obtained with original MCTS codes, and standard deviation (SD) and relative standard deviation (RSD) of this set of codesSupplementary Table S- 5 : $F_3$ values obtained with original MCTS codes, and standard deviation (SD) and relative standard deviation (RSD) of this set of codes) were also calculated and confirmed the same trend with a high relative standard deviation value of 104% for 50 eV, and again a decrease to 13% for 100 eV electrons. In this case, it is the sum of $P(0)$, $P(1)$ and $P(2)$ that is at the base of these values.

## Individual MCTS codes with common-set cross sections: dispersion in ICSDs

In this section, the results obtained with the modified track-structure codes are shown. As explained in the materials and methods section, these data were produced using the common-set interaction cross sections. Figure 3 shows the results of the ICSDs for monoenergetic electrons from 20 eV to 1 keV in spherical targets of 8 nm diameter. Figure 4 shows the same results for 5 and 10 keV electrons in larger target volumes of 100 nm diameter.

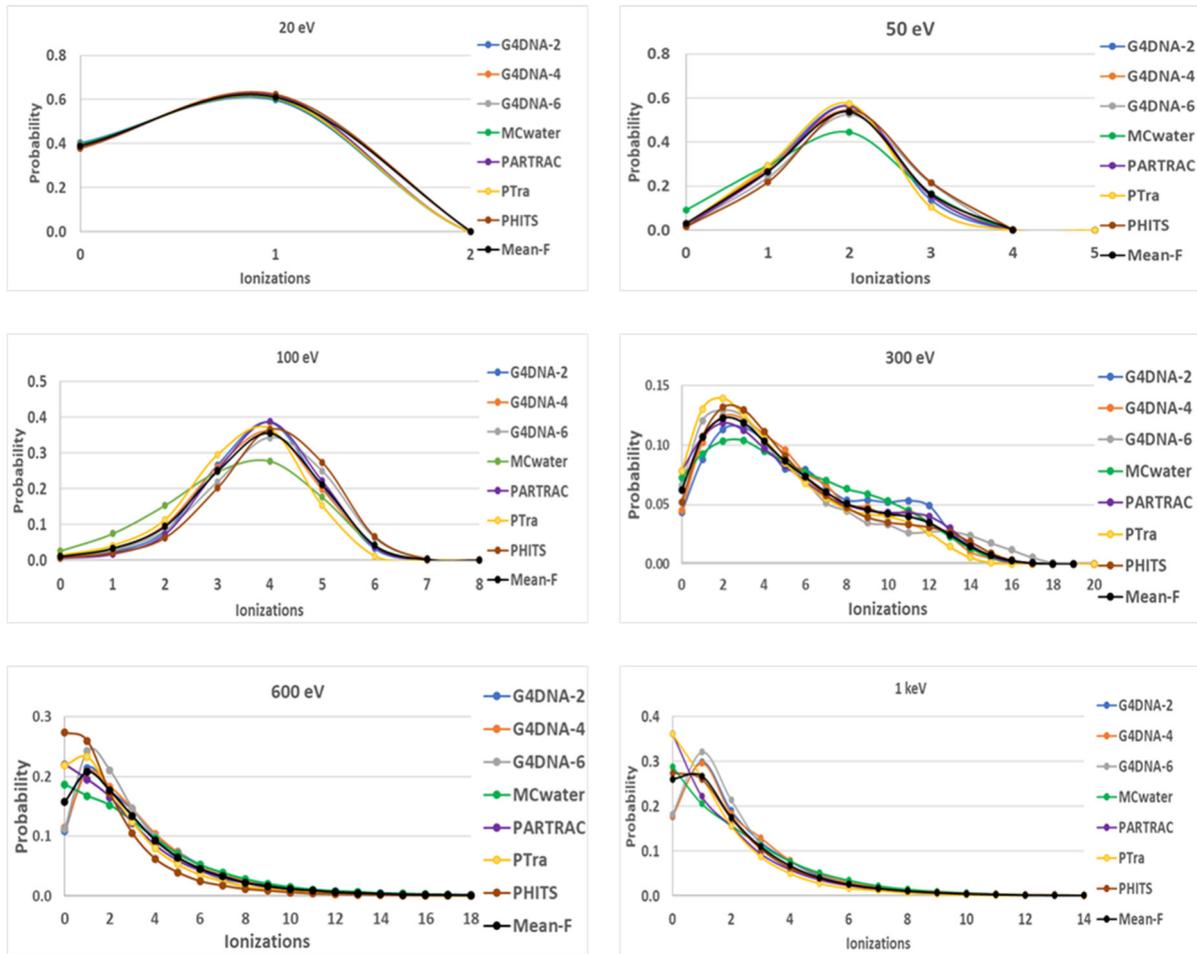

*Figure 3 : ICSDs predicted by diverse MCTS codes using the same common-set cross sections. The labels indicate starting electron energies in spherical targets of 8 nm diameter. The data shown in black are the ICSDs ("Mean-F") calculated as a mean frequency distribution of all the modified codes' ICSDs. Note: the lines are only there to guide the eye, as non-integer values for the number of ionizations are meaningless.*

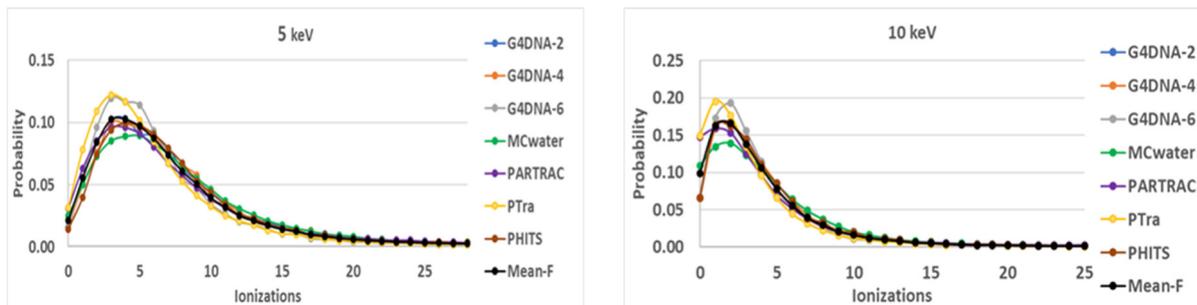

*Figure 4 : ICSDs obtained after modifying the MCTS codes with common cross sections for monoenergetic electrons of 5 and 10 keV in 100 nm diameter spherical targets. The data shown in black are the ICSDs ("Mean-F") calculated as a mean frequency distribution of all the modified codes' ICSDs. Note: the lines are only there to guide the eye, as non-integer values for the number of ionizations are meaningless.*

A comparison of Figures 3 and 4 with Figures 1 and 2 shows that the use of the common cross section dataset by all MCTS codes has notably reduced the dispersion of their predictions, in particular for low energy electrons. To quantify the residual dispersion of the results using the common cross sections , the corresponding Wasserstein-1 distances were calculated for each ICSD obtained with the modified codes to the mean ones shown in black colour in

Figure 3 and Figure 4. The results can be found in Supplementary Table S- 3 : Wasserstein distances calculated between each ICSD obtained with modified MCTS codes and the mean ICSD presented in black color in

Figure 3 and Figure 4. Table 5 shows the mean values of these Wasserstein-1 distances between the results of all modified codes and the mean distribution. As expected, the resulting values are dramatically lower than the ones obtained with original MCTS codes in Table 2 Table 2 : Mean value of the Wasserstein-1 distances between the ICSDs from the original codes with respect to the Mean ICSD for each electron energy. for electron energy of 20 eV where only 0 or 1 ionizations are possible in the target volume. This is because the target volume is large enough to contain 1 ionization irrespective of the angular distribution of the elastic scattering interactions. However, as the electron energy increases, the mean value of the distances increases and is about as high as that obtained with the original codes for 600 eV and higher energies.

*Table 5 : Mean value of the Wasserstein-1 distance between the ICSDs from the modified codes and the mean ICSD for each electron energy.*

| Energy (eV) | 20 | 50 | 100 | 300 | 600 | 1000 | 5000 | 10000 |
|---|---|---|---|---|---|---|---|---|
| Mean value of the W-1 distances for the modified codes | 0.007 | 0.08 | 0.18 | 0.32 | 0.37 | 0.27 | 0.62 | 0.40 |

## Individual MCTS codes with common-set cross sections: Dispersion in nanodosimetric quantities

Table 6 and Table 7 present the mean numbers of ionizations and probabilities to induce two or more ionizations (nanodosimetric quantities $M_1$ and $F_2$), respectively, with the individual MCTS codes using the common cross section data. They also show the relative standard deviation of these values for each electron energy. The reduced dispersion of ICSDs predicted by individual codes upon using the common cross section dataset, shown in Figures 3 and 4, leads to very low values of the relative standard deviation of the $M_1$ and $F_2$ values for the 20 eV and 50 eV electron sources (from 0.45 to 0.01 at 20 eV and from 0.23 to 0.05 at 50 eV). This decrease in the relative standard deviation of $M_1$ value with respect to the one shown initially in Table 3 is also observed for 100 eV and 300 eV but tends to disappear for electron energies of 600 eV and higher, where relative standard deviations of $M_1$ with the modified MCTS codes show similar values to those with the original codes.

Concerning $F_2$ values, the decrease in the relative standard deviation is large at 50 eV (from 0.34 to 0.04), but it doesn't show much variation for other electron energies, as it was already very low with the original MCTS codes (see Table 4 ). The same trend holds for $F_3$ values (Supplementary Table S- 6 : $F_3$ values obtained with modified MCTS codes using the common cross section and standard deviation (SD) and relative standard deviation (RSD) of this set of codes. ) where the RSD at 50 eV dramatically decreases from 1.04 to 0.25 with the modified codes but show similar values for the other electron energies.

Table 6 : Values of the first momentum (mean value) of the ICSDs calculated with the modified MCTS. The standard deviation (SD) and its relative value (RSD) for each electron energy are also given to quantify the dispersion of results.

| $M_1$: Mean numbers of ionizations predicted by the modified MCTS codes | | | | | | | | | |
|---|---|---|---|---|---|---|---|---|---|
| Energy (eV) | G4-DNA Opt 2 | G4-DNA Opt 4 | G4-DNA Opt 6 | PARTRAC | MCwater | PTra | PHITS | SD | RSD |
| 20 | 0.60 | 0.61 | 0.61 | 0.62 | 0.60 | 0.62 | 0.61 | 0.01 | 0.01 |
| 50 | 1.82 | 1.84 | 1.94 | 1.86 | 1.70 | 1.76 | 1.97 | 0.10 | 0.05 |
| 100 | 3.75 | 3.70 | 3.88 | 3.83 | 3.38 | 3.47 | 4.00 | 0.22 | 0.06 |
| 300 | 5.71 | 5.26 | 5.24 | 5.26 | 5.47 | 4.58 | 5.15 | 0.35 | 0.07 |
| 600 | 3.37 | 3.23 | 2.84 | 2.86 | 3.38 | 2.50 | 3.64 | 0.40 | 0.13 |
| 1000 | 2.31 | 2.29 | 2.04 | 1.79 | 2.29 | 1.58 | 2.11 | 0.28 | 0.14 |
| 5000 | 6.91 | 7.00 | 6.16 | 6.76 | 7.32 | 5.96 | 6.92 | 0.49 | 0.07 |
| 10000 | 4.26 | 4.26 | 3.84 | 3.96 | 4.51 | 3.38 | 4.05 | 0.37 | 0.09 |

Table 7 : Values for the cumulated probability of obtaining 2 or more ionizations ($F_2$) in the target volume calculated with the modified MCTS, as described in the materials and methods section. The standard deviation and its relative value for each of electron energy are also given to quantify the dispersion of results

| $F_2$: Probability of two or more ionizations predicted by the modified MCTS codes | | | | | | | | | |
|---|---|---|---|---|---|---|---|---|---|
| Enery (eV) | G4-DNA Opt 2 | G4-DNA Opt 4 | G4-DNA Opt 6 | PARTRAC | MCwater | PTra | PHITS | SD | RSD |
| 50 | 0.70 | 0.70 | 0.74 | 0.72 | 0.72 | 0.68 | 0.77 | 0.03 | 0.04 |
| 100 | 0.97 | 0.96 | 0.96 | 0.98 | 0.90 | 0.94 | 0.97 | 0.03 | 0.03 |
| 300 | 0.87 | 0.85 | 0.81 | 0.82 | 0.83 | 0.79 | 0.84 | 0.03 | 0.03 |
| 600 | 0.68 | 0.68 | 0.64 | 0.58 | 0.65 | 0.58 | 0.66 | 0.04 | 0.06 |
| 1000 | 0.52 | 0.53 | 0.50 | 0.42 | 0.51 | 0.37 | 0.47 | 0.06 | 0.12 |
| 5000 | 0.93 | 0.93 | 0.93 | 0.91 | 0.89 | 0.87 | 0.95 | 0.03 | 0.03 |
| 10000 | 0.77 | 0.78 | 0.76 | 0.69 | 0.74 | 0.65 | 0.76 | 0.05 | 0.06 |

### Potential impact of cross section differences on predicted initial DNA damage

In the simulations with PARTRAC for electrons started isotropically from homogeneously distributed points within the model cell nucleus, the average energy deposited to the nucleus by a 100 eV electron amounted to 99.95 eV using the original cross section database. That is, on average only 0.05% of the initial electron energy was deposited outside the nucleus. For 10 keV electrons, on average 8.39 keV were imparted in the nucleus, and on average about 16% of the initial electron energy was carried away from the nucleus. This is due to electrons started close to the surface of the nucleus which may travel and deposit a part of their energy outside the nucleus; their fraction is negligible for extremely short-ranged 100 eV electrons but notable at 10 keV with a path length of 2.77 μm and range as penetration depth of 1.41 μm. Using the common cross section database, these figures changed only marginally to 99.97 eV or 8.33 keV deposited to the nucleus per 100 eV or 10 keV electron, respectively.

To achieve a good statistic for the yields of DNA lesions, in total ten million tracks were simulated for 100 eV electrons and one million of tracks for 10 keV electrons, sub-divided into 100 runs in both cases. The doses corresponding to these numbers of primary electrons amounted in total to about 305.8 and 305.9 Gy (about 3.1 Gy per run) for the 100 eV electrons and the original and modified cross sections, respectively. For the 10 keV electrons, doses of 2650 and 2550 Gy (about 26 Gy per run) were obtained.

For 100 eV electrons, the predicted average yields of SSB amounted to 153.8 ± 0.3 per Gy per Gbp (mean ± standard error of the mean estimated from the 100 runs simulated) using the original cross section data. Upon replacing it with the common-set cross sections, the SSB yield decreased to 144.8 ± 0.3 per Gy per Gbp, i.e. by 5.9%. The average yields of DSB increased by 14.8% from 5.40 ± 0.05 to 6.20 ± 0.05 per Gy per Gbp upon replacing the original cross sections with the common ones. Only very rarely, a DSB cluster was scored at this low electron energy; the yields of DSB clusters increased from 0.003 ± 0.002 to 0.011 ± 0.002 per Gy per Gbp upon using the common set instead of original cross section data. None of these clusters included more than two DSBs.

For 10 keV electrons, upon replacing the original cross sections with the common-set ones, the predicted yields of SSB decreased by 2.5% from 151.6 ± 0.2 to 147.8 ± 0.2 per Gy per Gbp, the yields of DSB increased by 0.6% from 8.87 ± 0.02 to 8.92 ± 0.02 per Gy per Gbp, and the yields of DSB clusters increased by 7% from 0.100 ± 0.002 to 0.107 ± 0.002 per Gy per Gbp. For both cross section datasets, the DSB clusters predominantly (in more than 98% of cases) included just two DSBs per cluster.

## Discussion

Comparison of the ICSDs obtained with a representative panel of MCTS codes currently used in nanodosimetry and radiobiology modelling reveals a considerable dispersion among the codes. We have quantified it using a specific metric (Wasserstein-1 distance) and the relative standard deviations of the nanodosimetric quantities $M_1$, $F_2$ and $F_3$.

ICSDs are generally very different for very low-energy electrons, and this difference decreases with increasing electron energy, as shown by the values presented in Table 2. Of course, these differences also depend on the volume in which the ICSDs are calculated, as shown by the 5 keV versus 1 keV values in Table 2. As the ICSDs for 5 keV electrons were calculated in a much larger volume than those for 1 keV (100 nm diameter versus 8 nm), the part of the track taken into account is larger and incorporates ultimately lower-energy interactions.

On the other hand, nanodosimetric quantities $M_1$, $F_2$ and $F_3$ are much less sensitive to the different MCTS codes, particularly at energies of around 100 or 300 eV. As these quantities are used in various models (Conte et al. 2018) to establish a link between the physical characteristics of the tracks and their biological impact at cellular level, these results show that the choice of MCTS code for this type of calculation may be less influential than the evaluation of the ICSD as a whole might suggest.

Nevertheless, the aim of this work was not simply to show and quantify these differences based on a panel of MCTS codes, which is certainly representative, but not exhaustive. Instead, the aim was to elucidate and quantify the contribution of the use of different interaction sections to these differences. Thus, the participants in this work modified the codes to use the same probability of achieving ionization or electronic excitation or elastic interaction with liquid water for a given electron energy. The kinetic energies of the secondary electrons and the angular distributions after these interactions were left unchanged from the original versions of the codes.

The relative differences between the initial results and those obtained with the modified codes allowed assessing the contribution of the choice of interaction cross section dataset to the initial dispersion of the results. This contribution was found to be dominant for low-energy electrons (96% for 20 eV electrons and 67% at 100 eV as can be calculated from the relative differences between values in Table 2 and Table 5). However, this contribution decreases with increasing electron energy, as the other elements of the simulation codes remain unchanged and have an influence on the shape of the tracks, which becomes increasingly important with growing electron energy. The volume over which the ICSDs are calculated also has an impact on the results, as it determines the portion of the track to be analyzed. Thus, the larger the volume, the larger the part of track analyzed. If the track portion analyzed is very large or even complete, then the number of ionizations produced will tend to move closer together between codes, because the codes use the same interaction cross sections. Under these conditions, the influence of the track shape (dominated by differential elastic scattering cross sections) and the secondary electron spectrum will tend to diminish. On the other hand, the influence of differential ionization and elastic cross sections is predominant for small volumes where only a portion of the track is captured in the ICSD (notably at the beginning of the track).

With regard to nanodosimetric quantities $M_1$, $F_2$ and $F_3$, it was found that the use of the same interaction cross sections by the different codes virtually canceled their dispersion. At the energy range analyzed here (20 eV-10 keV), the remaining dispersion is at the level of the statistical uncertainties, with a possible exception of energies around 1 keV. In this work, the tracks of 1 keV electrons were analyzed in the small 8 nm diameter volume, where the angular dependence of the elastic cross sections plays a very important role for the number of ionizations finally counted in the volume. In more tangible terms, it can therefore be stated that interaction cross sections are mainly responsible for the differences found in the nanodosimetric quantities calculated from ICSDs in volumes corresponding in size to those of nucleosomes or the occurrence of complex DNA damage. For electrons from ~1keV upwards, the angular distribution of elastic interactions and the differential ionization cross sections can also impact nanodosimetric quantities by around 10-15%.

To corroborate these conclusions, the approach taken in this study was carried one step further exploiting the ease of changing cross sections and models in the Geant4-DNA code. Additional simulations were performed with the three options (2, 4 and 6) using the common cross section data set as previously and using in addition the same elastic differential cross sections (those of option 2, i.e., Champion's model, see Table 1). Thus, only the differential ionization cross sections and the resulting angular distributions of the electrons after an ionization remained different between the options.

The results are shown in Supplementary Figure S-4. What can be observed is that options 2 and 4 are now giving almost identical results, which is understandable given that the inelastic collision models used are very similar (see Table 1). Option 6, on the other hand, still gives ICSDs that are visibly different from those of the other two options. However, the values of the mean of the distributions, $M_1$, for example, remain virtually unchanged from the values presented in Table 6.

This conclusion is in line with the one found by Emfieztzoglou and collaborators [Emfietzoglou et al. 2011] where they demonstrated that the choice of the extension algorithm used to extrapolate optical data to finite momentum transfer in order to calculate differential inelastic cross-sections for each ionization shell was crucial in modelling the inelastic scattering of electrons with energies below 200 eV.

It is important to point out once again that In view of the great difficulty of determining which are the correct values of the electron cross sections, from a purely theoretical point of view, one could state that the average values of the different cross sections available are those that come closest to this "true value". However, in this exercise, only a representative but not exhaustive panel of existing values was included, and so the common cross section data set circulated to modify the codes cannot be considered as an approximation to the true values of the cross sections, even from a theoretical point of view.

*Having said that, we nevertheless wanted to assess how the ICSDs of each of the codes had changed with the common-set of interaction cross sections. Therefore, in*

Supplementary Table S- 4 : Wasserstein distances calculated between each ICSD obtained with the original MCTS code and its modified version using common cross sections used in this work), values of Wasserstein-1 distances are presented between the results obtained with the original versions of the codes and their modified versions. It can be seen, for example, that large Wasserstein-1 values are obtained for the PHITS code original-versus the modified version. That stems from the fact that in the original version, electronic excitations were followed by autoionization (90% for the diffuse band and 100% for collective excitations), whereas this is not the case in the modified version.

The reported yields of DNA damage simulations obtained with the PARTRAC code are in line with previous studies assessing biological effects of diverse types of radiations (Friedland et al 2017, 2019, Kundrát et al 2020). The reported results indicate that the use of different cross section datasets may lead to considerable differences in predicted DNA damage yields. At both studied electron energies, 100 eV and 10 keV, there is a clear trend towards alterations in damage clustering. Original PARTRAC cross sections assume much less inelastic interactions at very low electron energies (up to about 100 eV) than the common-set ones (Supplementary Figure S-1). Consequently, for 100 eV electrons, the predicted DNA strand breaks for the common-set cross sections more frequently appear closer to each other, so that they are less frequently isolated as SSB (-6%) and more often cluster into a DSB (+15%) as compared to original PARTRAC cross sections. For 10 keV electrons, the differences are not as pronounced; they manifest as an approximately 2.5% reduction in SSB yields and 0.6% increase in DSB numbers. The predicted impact on the yields of DSB clusters is even larger at both energies, but these complex lesions remain very rare for low-energy electrons anyway.

The two studied energies are relevant for beta decaying isotopes such as $^3$H whose mean electron energy is about 6 keV, and for Auger emitters such as $^{125}$I that produces electrons with energies including the 50-500 eV range. Interestingly, some works clearly show that the significant differences in the final number of DSBs simulated with different MCTS codes are not only the consequence of differences in the patterns of the charged particle tracks. Indeed, other elements are just as important, or even much more influential such as: the DNA geometries used (Tang et al. 2029, Thibaut et al. 2022), the energy threshold criteria for producing a direct or indirect break (Thompson et al 2022, Chatzipapas et al 2024), and the very definition of a DSB or its complexity (Schuemann et al. 2019). These parameters can result in a factor of more than two in the number of DSBs simulated by these methods (Le Tuhan et al. 2024). However, for a given simulation code, having fixed these factors, the reported DNA damage simulations performed with the PARTRAC code give an illustration on how DSB yields can vary with different cross section data and show that, especially for low energy electrons, the cross section-related variability may easily reach as much as 15% for DSBs and may be even higher for more complex lesions such as DSB clusters. Given the critical impact of these lesions in cell killing by ionizing radiation and the role electrons play in interactions of diverse radiation types including X- or γ-rays but also protons and ions, interaction cross sections of low-energy electrons in liquid water and biologically relevant media do deserve further research on both the experimental and theoretical side.

## Conclusion

The present analysis underscores the critical role of interaction cross sections for nanodosimetric predictions. Variability in cross section datasets across MCTS codes significantly affects ICSDs and derived nanodosimetric quantities, especially for low-energy electrons. By harmonizing cross section datasets, this study reduced the dispersion of simulation results, demonstrating that cross sections account for up to 96% of ICSD variability at 20 eV and 67% at 100 eV. The study also shows that derived nanodosimetric quantities such as $M_1$ and $F_2$ exhibit lower variability than ICSDs. However, discrepancies persist at higher electron energies due to other factors, including angular distributions and secondary electron spectra. This suggests that while total cross section data are preeminent at low energies, other factors gain importance with increasing electron energy.

DNA damage simulations using PARTRAC illustrated the biological significance of cross section variability. Differences in predicted SSB and DSB yields showed indications that low-energy electron interactions can affect damage clustering, with implications for understanding radiation-induced cell inactivation. The reported variability in DSB predictions, reaching up to 15%, underscores the need for precision in cross section modeling, particularly for radiation types producing low-energy electrons, such as Auger emitters.

Ultimately, as single shell interaction cross section data for low-energy electrons in condensed matter cannot be directly validated experimentally, this work highlights the need for high-quality experimental data on integrated physical values derived from the electron tracks or even biological endpoints such as DSBs. Such data would provide a more direct basis for comparing

MCTS codes and improving their reliability. Such advancements are pivotal for improving the predictive power of MCTS codes in radiobiology, radiation protection, and medical applications.

# References

[Allison et al. 2006] Allison J, et al. Geant4 developments and applications. IEEE Trans Nucl Sci. 2006;53:270-278. DOI:10.1109/TNS.2006.869826.

[Allison et al. 2016] Allison J, et al. Recent developments in Geant4. Nucl Instrum Meth A. 2016;835:186-225. DOI:10.1016/j.nima.2016.06.125.

[Agostinelli et al. 2003] Agostinelli S, et al. Geant4 - A simulation toolkit. Nucl Instrum Meth A. 2003;506:250-303. DOI:10.1016/S0168-9002(03)01368-8.

[Bantsar et al. 2018] Bantsar A, Colautti P, Conte V, Hilgers G, Pietrzak M, Pszona S, Rabus H, Selva A. State of the art of instrumentation in experimental nanodosimetry. Radiat Prot Dosim. 2018;180:177-181. DOI:10.1093/rpd/ncx263.

[Berger et al. 1993] Berger MJ, Seltzer SM, Wang R, Schechter A. Elastic scattering of electrons and positrons by atoms: database ELAST. NISTIR 5166, U.S. Department of Commerce, Gaithersburg, MD; 1993.

[Bernal et al. 2015] Bernal MA, Bordage MC, Brown JMC, Davídková M, Delage E, El Bitar Z, Enger SA, Francis Z, Guatelli S, Ivanchenko VN, Karamitros M, Kyriakou I, Maigne L, Meylan S, Murakami K, Okada S, Payno H, Perrot Y, Petrovic I, Pham QT, Ristic-Fira A, Sasaki T, Štěpán V, Tran HN, Villagrasa C, Incerti S. Track structure modeling in liquid water: a review of the Geant4-DNA very low energy extension of the Geant4 Monte Carlo simulation toolkit. Phys Medica. 2015;31:861-874. DOI:10.1016/j.ejmp.2015.10.087.

[Bigildeev and Michalik 1996] Bigildeev EA, Michalik V. Charged particle tracks in water of different phases: Monte Carlo simulation of electron tracks. Radiat Phys Chem. 1996;47:197-207. DOI:10.1016/0969-806X(95)00002-F.

[Bordage et al. 2016] Bordage M, Bordes J, Edel S, Terrissol M, Franceries X, Bardiès M, Lampe N, Incerti S. Implementation of new physics models for low energy electrons in liquid water in Geant4-DNA. Phys Medica. 2016;32:1833-1840. DOI:10.1016/j.ejmp.2016.10.006.

[Bug et al. 2013] Bug M, Gargioni E, Nettelbeck H, Baek WY, Hilgers G, Rozenfeld A, Rabus H. Ionization cross section data of nitrogen, methane, and propane for light ions and electrons and their suitability for use in track structure simulations. Phys Rev E. 2013;88:043308. DOI:10.1103/PhysRevE.88.043308.

[Bug et al. 2017] Bug M, Baek WY, Rabus H, Villagrasa C, Meylan S, Rosenfeld AB. An electron-impact cross section data set (10 eV–1 keV) of DNA constituents based on consistent experimental data: a requisite for Monte Carlo simulation set. Radiat Phys Chem. 2017;130:459-479. DOI:10.1016/j.radphyschem.2016.09.027.

[Brenner and Zaider 1983] Brenner DJ, Zaider M. Phys Med Biol. 1983;29:443-447.

[Champion et al. 2009] Champion C, Incerti S, Aouchiche H, Oubaziz D. A free-parameter theoretical model for describing the electron elastic scattering in water in the Geant4 toolkit. Radiat Phys Chem. 2009;78:745-750. DOI:10.1016/j.radphyschem.2009.03.079.

[Champion et al. 2014] Champion C, Quinto MA, Bug MU, Baek WY, Weck PF. Theoretical and experimental quantification of doubly and singly differential cross sections for electron-induced ionization of isolated tetrahydrofuran molecules. Eur Phys J D. 2014;68:205. DOI:10.1140/epjd/e2014-40829-8.

[Chatzipapas et al. 2024] Chatzipapas KP, Tran NH, Dordevic M, Sakata D, Incerti S, Visvikis D, Bert J. Development of a novel computational technique to create DNA and cell geometrical models for Geant4-DNA. Phys Medica. 2024;127:104839. DOI:10.1016/j.ejmp.2024.104839.

[Conte et al. 2023] Conte V, Bianchi A, Selva A. Track structure of light ions: the link to radiobiology. Int J Mol Sci. 2023;24(6):5826. DOI:10.3390/ijms24065826.

[Conte et al. 2018] Conte V, Selva A, Colautti P, Hilgers G, Rabus H, Bantsar A, Pietrzak M, Pszona S. Nanodosimetry: towards a new concept of radiation quality. Radiat Prot Dosim. 2018;180:1-4, 150-156. DOI:10.1093/rpd/ncx175.

[Cuccinotta et al. 2003] Cuccinotta H, Francis A, Wu S, Shavers MR, George K. Radiation dosimetry and biophysical models of space radiation effects. Gravit Space Biol. 2003;16:11-18.

[Dingfelder et al. 1998] Dingfelder M, Hantke D, Inokuti M, Paretzke HG. Electron inelastic-scattering cross sections in liquid water. Radiat Phys Chem. 1998;53:1-18. DOI:10.1016/S0969-806X(97)00317-4.

[Dingfelder et al. 2008] Dingfelder M, Ritchie RH, Turner JE, Friedland W, Paretzke HG, Hamm RN. Comparisons of calculations with PARTRAC and NOREC: transport of electrons in liquid water. Radiat Res. 2008;169:584-594. DOI:10.1667/RR1099.1.

[Dingfelder et al. 2014] Dingfelder M. Updated model for dielectric response function of liquid water. Appl Radiat Isot. 2014;83:142-147. DOI:10.1016/j.apradiso.2013.01.016.

[Durante et al. 2021] Durante M, Debus J, Loeffler J. Physics and biomedical challenges of cancer therapy with accelerated heavy ions. Nat Rev Phys. 2021;3:1-14. DOI:10.1038/s42254-021-00368-5.

[Edwards 2001] Edwards AA. RBE of radiation in space and the implications for space travel. Phys Medica. 2001;17:147-152.

[Emfietzoglou et al. 2011] Emfietzoglou D, Kyriakou I, Abril I, Garcia-Molina R, Nikjoo H. Inelastic scattering of low-energy electrons in liquid water computed from optical-data models of the Bethe surface. Int J Radiat Biol. 2011;88(1-2):22-28. DOI:10.3109/09553002.2011.588061.

[Faddegon et al. 2020] Faddegon B, Ramos-Mendez J, Schuemann J, McNamara A, Shin J, Perl J, Paganetti H. The TOPAS tool for particle simulation: a Monte Carlo simulation tool for physics, biology and clinical research. Phys Medica. 2020;72:114-121. DOI:10.1016/j.ejmp.2020.03.019.

[Friedland et al. 2011a] Friedland W, Dingfelder M, Kundrat P, Jacob P. Track structures, DNA targets, and radiation effects in the biophysical Monte Carlo simulation code PARTRAC. Mutat Res. 2011;711:28-40. DOI:10.1016/j.mrfmmm.2011.01.003.

[Friedland et al. 2011b] Friedland W, Jacob P, Kundrat P. Mechanistic simulation of radiation damage to DNA and its repair: on the track towards systems radiation biology modelling. Radiat Prot Dosim. 2011;143:542-548. DOI:10.1093/rpd/ncq383.

[Friedland et al. 2017] Friedland W, Schmitt E, Kundrat P, et al. Comprehensive track-structure-based evaluation of DNA damage by light ions from radiotherapy-relevant energies down to stopping. Sci Rep. 2017;7:45161. DOI:10.1038/srep45161.

[Friedland et al. 2019] Friedland W, Kundrat P, Schmitt E, Becker J, Li W. Modeling DNA damage by photons and light ions over energy ranges used in medical applications. Radiat Prot Dosim. 2019;183(1-2):84-88. DOI:10.1093/rpd/ncy245.

[Givens et al. 1984] Givens CR, Shortt RM. A class of Wasserstein metrics for probability distributions. Mich Math J. 1984;31(2):231-240. DOI:10.1307/mmj/1029003026.

[Grosswendt 2002] Grosswendt B. Formation of ionization clusters in nanometric structures of propane-based tissue-equivalent gas or liquid water by electrons and alpha-particles. Radiat Environ Biophys. 2002;41:103-112. DOI:10.1007/s00411-002-0155-6.

[Grosswendt and Waibel 1978] Grosswendt B, Waibel E. Transport of low energy electrons in nitrogen and air. Nucl Instrum Meth. 1978;155:145-156. DOI:10.1016/0029-554X(78)90198-2.

[Grosswendt et al. 2004] Grosswendt B, De Nardo L, Colautti P, Pszona S, Conte V, Tornielli G. Experimental equivalent cluster-size distributions in nanometric volumes of liquid water. Radiat Prot Dosim. 2004;110(1-4):851-857. DOI:10.1093/rpd/nch203.

[Hayashi et al. 2000] Hayashi H, Watanabe N, Udagawa Y, Kao CC. The complete optical spectrum of liquid water measured by inelastic x-ray scattering. Proc Natl Acad Sci USA. 2000;97:6264-6266. DOI:10.1073/pnas.110572097.

[Heller et al. 1974] Heller JM, Hamm RN, Birkhoff RD, Painter LR. Collective oscillation in liquid water. J Chem Phys. 1974;60:3483-3486. DOI:10.1063/1.1681563.

[ICRU 1984] International Commission on Radiation Units and Measurements. Stopping powers for electrons and positrons. ICRU Report 37. Bethesda, MD: International Commission on Radiation Units and Measurements; 1984.

[ICRU 2014] International Commission on Radiation Units and Measurements. Key data for ionizing-radiation dosimetry: measurement standards and applications. ICRU Report 90. Volume 14, Issue 1; 2014.

[ICRP 2013] International Commission on Radiological Protection. Assessment of radiation exposure of astronauts in space. ICRP Publication 123. Ann ICRP. 2013;42(4).

[Incerti et al. 2010a] Incerti S, Baldacchino G, Bernal M, Capra R, Champion C, Francis Z, Guatelli S, Guèye P, Mantero A, Mascialino B, Moretto P, Nieminen P, Rosenfeld A, Villagrasa C, Zacharatou C. The Geant4-DNA project. Int J Model Simul Sci Comput. 2010;1:157-178. DOI:10.1142/S1793962310000122.

[Incerti et al. 2010b] Incerti S, Ivanchenko A, Karamitros M, Mantero A, Moretto P, Tran HN, Mascialino B, Champion C, Ivanchenko VN, Bernal MA, Francis Z, Villagrasa C, Baldacchino G, Guèye P, Capra R, Nieminen P, Zacharatou C. Comparison of Geant4 very low energy cross section models with experimental data in water. Med Phys. 2010;37:4692-4708. DOI:10.1118/1.3476457.

[Incerti et al. 2018] Incerti S, Kyriakou I, Bernal MA, Bordage MC, Francis Z, Guatelli S, Ivanchenko V, Karamitros M, Lampe N, Lee SB, Meylan S, Min CH, Shin WG, Nieminen P, Sakata D, Tang N, Villagrasa C, Tran HN, Brown JMC. Geant4-DNA example applications for track structure simulations in liquid water: a report from the Geant4-DNA Project. Med Phys. 2018;45:e722-e739. DOI:10.1002/mp.13048.

[Kai et al. 2024] Kai T, Toigawa T, Matsuya Y, et al. Significant role of secondary electrons in the formation of a multi-body chemical species spur produced by water radiolysis. Sci Rep. 2024;14:24722. DOI:10.1038/s41598-024-76481-z.

[Kaplan et al. 1990] Kaplan IG, Miterev AM, Sukhonosov VYa. Simulation of the primary stage of liquid water radiolysis. Radiat Phys Chem. 1990;36(3):493-498. DOI:10.1016/1359-0197(90)90039-K.

[Kim et al. 2000] Kim YK, Santos JP, Parente F. Extension of the binary-encounter-dipole model to relativistic incident electrons. Phys Rev A. 2000;62:052710. DOI:10.1103/PhysRevA.62.052710.

[Krämer and Scholz 2006] Krämer M, Scholz M. Rapid calculation of biological effects in ion radiotherapy. Phys Med Biol. 2006;51:1959-1970. DOI:10.1088/0031-9155/51/8/001.

[Kulesza et al. 2022] Kulesza JA, Adams TR, Armstrong JC, Bolding SR, Brown FB, Bull JS, Burke TP, Clark AR, Forster RA III, Giron JF, Grieve TS, Josey CJ, Martz RL, McKinney GW, Pearson EJ, Rising ME, Solomon CJ Jr, Swaminarayan S, Trahan TJ, Wilson SC, Zukaitis AJ. MCNP® Code Version 6.3.0 Theory & User Manual. Los Alamos National Laboratory Tech Rep. LA-UR-22-30006, Rev. 1, Los Alamos, NM, USA; 2022.

[Kundrat et al. 2020] Kundrat P, Friedland W, Becker J, et al. Analytical formulas representing track-structure simulations on DNA damage induced by protons and light ions at radiotherapy-relevant energies. Sci Rep. 2020;10:15775. DOI:10.1038/s41598-020-72857-z.

[Kyriakou et al. 2015] Kyriakou I, Incerti S, Francis Z. Improvements in Geant4 energy-loss model and the effect on low-energy electron transport in liquid water. Med Phys. 2015;42:3870-3876. DOI:10.1118/1.4921613.

[Kyriakou et al. 2022] Kyriakou I, Sakata D, Tran HN, Perrot Y, Shin WG, Lampe N, Zein S, Bordage MC, Guatelli S, Villagrasa C, Emfietzoglou D, Incerti S. Review of the Geant4-DNA simulation toolkit for radiobiological applications at the cellular and DNA level. Cancers. 2022;14:35. DOI:10.3390/cancers14010035.

[Kyriakou et al. 2025] Kyriakou I, Tran HN, Desorgher L, Ivantchenko V, Guatelli S, Santin G, Niemiene P, Incerti S, Emfietzoglou D. Extension of the discrete electron transport capabilities of the Geant4-DNA toolkit to MeV energies. Appl Sci. 2025;15:1183. DOI:10.3390/app15031183.

[Lazarakis et al. 2012] Lazarakis P, Bug MU, Gargioni E, Guatelli S, Rabus H, Rosenfeld AB. Phys Med Biol. 2012;57:1231. DOI:10.1088/0031-9155/57/5/1231.

[Le Tuan et al. 2024] Le Tuan A, Tran HN, Thibaut Y, Chatzipapas K, Sakata D, et al. "Dsbandrepair": An updated Geant4-DNA simulation tool for evaluating the radiation-induced DNA damage and its repair. Phys Medica. 2024;124:103422. DOI:10.1016/j.ejmp.2024.103422.

[Li et al. 2018] Li WB, Hofmann W, Friedland W. Microdosimetry and nanodosimetry for internal emitters. Radiat Meas. 2018;115:29-42. DOI:10.1016/j.radmeas.2018.05.013.

[Matsuya et al. 2019] Matsuya Y, Kai T, Yoshii Y, Yachi Y, Naijo S, Date H, Sato T. Modeling of yield estimation for DNA strand breaks based on Monte Carlo simulations of electron track structure in liquid water. J Appl Phys. 2019;126(12):124701. DOI:10.1063/1.5115519.

[Matsuya et al. 2022] Matsuya Y, Kai T, Sato T, Ogawa T, Hirata Y, Yoshii Y, Parisi A, Liamsuwan T. Track-structure modes in particle and heavy ion transport code system (PHITS): Application to radiobiological research. Int J Radiat Biol. 2022;98(2):148-157. DOI:10.1080/09553002.2022.2013572.

[Meylan et al. 2017] Meylan S, Incerti S, Karamitros M, Tang N, Bueno M, Clairand I, et al. Simulation of early DNA damage after the irradiation of a fibroblast cell nucleus using Geant4-DNA. Sci Rep. 2017;7:11923. DOI:10.1038/s41598-017-11851-4.

[Michaud and Sanche 1987] Michaud M, Sanche L. Absolute vibrational excitation cross sections for slow-electron (1-18 eV) scattering in solid H2O. Phys Rev A Gen Phys. 1987;36(10):4684-4699. DOI:10.1103/physreva.36.4684.

[Moliere 1948] Moliere G. Theorie der Streuung schneller geladener Teilchen II: Mehrfach- und Vielfachstreuung. Z Naturforsch A. 1948;3a:78-97. DOI:10.1515/zna-1948-0203.

[Neuefeind et al. 2002] Neuefeind J, Benmore CJ, Tomberli B, Egelstaff PA. Experimental determination of the electron density of liquid H2O and D2O. J Phys Condens Matter. 2002;14(23):L429. DOI:10.1088/0953-8984/14/23/104.

[Nikjoo et al. 1994] Nikjoo H, Charlton DE, Goodhead DT. Monte Carlo track structure studies of energy deposition and calculation of initial DSB and RBE. Adv Space Res. 1994;14:161-180. DOI:10.1016/0273-1177(94)90466-9.

[Paretzke 1988] Paretzke HG. Simulation von Elektronenspuren in Energiebereich 0.01–10 keV in Wasserdampf. GSF-Bericht. Institut fur Strahlenschutz der Gesellschaft fur Strahlen-und Umweltforschung; 1988. p. 1-87.

[Penn 1976] Penn DR. Electron mean free paths for free-electron-like materials. Phys Rev B. 1976;13(12):5248. DOI:10.1103/PhysRevB.13.5248.

[Penn 1987] Penn DR. Electron mean-free-path calculations using a model dielectric function. Phys Rev B. 1987;35(2):482. DOI:10.1103/PhysRevB.35.482.

[Pietrzak et al. 2021] Pietrzak M, Mietelska M, Bancer A, Rucinski A, Brzozowska B. Geant4-DNA modeling of nanodosimetric quantities in the Jet Counter for alpha particles. Phys Med Biol. 2021;66(22):225008. DOI:10.1088/1361-6560/ac33eb.

[Rabus et al. 2022] Rabus H, Zankl M, Gómez-Ros JM, Villagrasa C, Eakins J, Huet C, Brkic H, Tanner R. Lessons learnt from the recent EURADOS intercomparisons in computational dosimetry. Radiat Meas. 2022;156:106822. DOI:10.1016/j.radmeas.2022.106822.

[Salvat et al. 2005] Salvat F, Jablonski A, Powell CJ. ELSEPA—Dirac partial-wave calculation of elastic scattering of electrons and positrons by atoms, positive ions and molecules. Comput Phys Commun. 2005;165:157-190. DOI:10.1016/j.cpc.2020.107704.

[Sarrut et al. 2022] Sarrut D, Arbor N, Baudier T, Borys D, Etxebeste A, Fuchs H, Gajewski J, Grevillot L, Jan S, Kagadis GC, Kang HG, Kirov A, Kochebina O, Krzemien W, Lomax A, Papadimitroulas P, Pommranz C, Roncali E, Rucinski A, Winterhalter C, Maigne L. The Open

GATE ecosystem for Monte Carlo simulation in medical physics. Phys Med Biol. 2022;67:184001. DOI:10.1088/0031-9155/49/19/007.

[Sato et al. 2024] Sato T, Iwamoto Y, Hashimoto S, Ogawa T, Furuta T, Abe S, Kai T, Matsuya Y, Matsuda N, Hirata Y, Sekikawa T, Yao L, Tsai PE, Ratliff HN, Iwase H, Sakaki Y, Sugihara K, Shigyo N, Sihver L, Niita K. Recent improvements of the Particle and Heavy Ion Transport code System - PHITS version 3.33. J Nucl Sci Technol. 2024;61:127-135. DOI:10.1080/00223131.2023.2275736.

[Schulte et al. 2008] Schulte RW, Wroe AJ, Bashkirov VA, Garty GY, Breskin A, Chechik R, Shchemelinin S, Gargioni E, Grosswendt B, Rosenfeld AB. Nanodosimetry-based quality factors for radiation protection in space. Z Med Phys. 2008;18:286-296. DOI:10.1016/j.zemedi.2008.06.011.

[Shin et al. 2021] Shin W, Sakata D, Lampe N, Belov O, Tran NH, et al. A Geant4-DNA evaluation of radiation-induced DNA damage on a human fibroblast. Cancer. 2021;3:4940. DOI:10.3390/cancers13194940.

[Shinotsuka et al. 2017] Shinotsuka H, Da B, Tanuma S, Yoshikawa H, Powell CJ, Penn DR. Calculations of electron inelastic mean free paths. XI. Data for liquid water for energies from 50 eV to 30 keV. Surf Interface Anal. 2017;49(4):238-252. DOI:10.1002/sia.6123.

[Scholz et al. 2020] Scholz M, Friedrich T, Magrin G, Colautti P, Ristic-Fira A, Petrovic I. Characterizing radiation effectiveness in ion beam therapy part I: introduction and biophysical modeling of RBE using the LEM IV. Front Phys. 2020;8:272. DOI:10.3389/fphy.2020.00272.

[Schuemann et al. 2019] Schuemann J, McNamara A, Warmenhoven J, et al. A new standard DNA damage (SDD) data format. Radiat Res. 2019;191:76-92. DOI:10.1667/RR15209.1.

[Swadia et al. 2017] Swadia M, Thakar Y, Vinodkumar M, Limbachiya C. Theoretical electron impact total cross sections for tetrahydrofuran (C4H8O). Eur Phys J D. 2017;71:85. DOI:10.1140/epjd/e2017-70617-9.

[Tang et al. 2019] Tang N, Bueno M, Meylan S, Incerti S, Tran H, Vaurijoux A, et al. Influence of chromatin compaction on simulated early radiation-induced DNA damage using Geant4-DNA. Med Phys. 2019;46:1501-1511. DOI:10.1002/mp.13405.

[Thibaut et al. 2022] Thibaut Y, Tang N, Tran HN, Vaurelie A, Villagrasa C, Perrot Y. Nanodosimetric calculations of radiation-induced DNA damage in a new nucleus geometrical model based on the isochore theory. Int J Mol Sci. 2022;23:3770. DOI:10.3390/ijms23073770.

[Tran et al. 2024] Tran HN, Archer, Baldacchino JG, Brown JMC, Chappuis F, Cirrone GAP, Desorgher L, Dominguez N, Fattori S, Guatelli S, Ivantchenko V, Méndez JR, Nieminen P, Perrot Y, Sakata D, Santin G, Shin WG, Villagrasa C, Zein S, Incerti S. Review of chemical models and applications in Geant4-DNA: report from the ESA BioRad III Project. Med Phys. 2024;5:5873-5889. DOI:10.1002/mp.17256.

[Uehara et al. 1992] Uehara S, Nikjoo H, Goodhead DT. Cross sections for water vapour for the Monte Carlo electron track structure code from 10 eV to the MeV region. Phys Med Biol. 1992;37:1841-1858. DOI:10.1088/0031-9155/38/12/010.

[Villagrasa et al. 2022] Villagrasa C, Rabus H, Baiocco G, Perrot Y, Parisi A, Struelens L, Qiu R, Beuve M, Poignant F, Pietrzak M, Nettelbeck H. Intercomparison of micro- and nanodosimetry Monte Carlo simulations: an approach to assess the influence of different cross sections for low-energy electrons on the dispersion of results. Radiat Meas. 2022;150:106675. DOI:10.1016/j.radmeas.2021.106675.

[Zaider 1996] Zaider M. Microdosimetric-based risk factors for radiation received in space activities during a trip to Mars. Health Phys. 1996;70:845-851.

[Zein et al. 2023] Zein SA, Bordage MC, Tran HN, Macetti G, Genoni A, Dal Cappello C, Incerti C. Simulation of low-energy ion tracks in water with a new Geant4-DNA toolkit version. Nucl Instrum Meth B. 2023;542:51-60. DOI:10.1016/j.nimb.2023.06.004.

# Supplementary materials

## Supplement 1 : Wasserstein-1 distance for ICSDs differences quantification:

In the variety of metrics that have been proposed in mathematical statistics for measuring distances between probability or frequency distribution, one that we believe to be conceptually simple yet useful for the present work is the Wasserstein $W_1$ distance (Givens and Scott 1984), defined for two discrete distributions as:

$$W = \sum_{i}^{bins}|\delta_i| \quad wih \quad \begin{cases} \delta_0 = 0 \\ \delta_{i+1} = \delta_i + x_i - y_i \end{cases} \quad \text{Eq(1)}$$

where $x_i$ corresponds to probability in the i-th bin of the first distribution and $y_i$ is the corresponding probability of the second distribution. $\delta_{i+1}$ is the difference between the cumulative distributions in the i-th bin. This metric is also called the Earth-Mover (EM) distance as it can be interpreted as the minimum energy cost (in physical terms: work) of moving and transforming bin by bin (pile by pile) one probability distribution into the shape of the other distribution. For instance, if one normalized distribution is obtained just by shifting another one, their Wasserstein-1 distance equals that shift. Therefore, the higher the Wasserstein-1 distance, the higher is the difference between the compared distributions for the same number of bins taken into account.

## /Supplementary Tables

## Participants data with original codes

*Supplementary Table S- 1: original participants data on ICSDs*

| Energy (eV) | ICS | G4-DNA Opt 2 | G4-DNA Opt 4 | G4-DNA Opt 6 | PARTRAC | MCwater | PTra | PHITS |
|---|---|---|---|---|---|---|---|---|
| 20 | 0 | 2.32E-01 | 5.63E-01 | 1.67E-01 | 7.92E-01 | 7.03E-01 | 3.78E-01 | 5.45E-01 |
| 20 | 1 | 7.68E-01 | 4.37E-01 | 8.33E-01 | 2.08E-01 | 2.97E-01 | 6.22E-01 | 4.55E-01 |
| 50 | 0 | 9.72E-03 | 1.61E-02 | 2.01E-03 | 1.15E-01 | 6.09E-02 | 1.22E-02 | 1.83E-03 |
| 50 | 1 | 1.01E-01 | 2.98E-01 | 6.48E-02 | 5.71E-01 | 4.99E-01 | 2.47E-01 | 1.78E-01 |
| 50 | 2 | 5.66E-01 | 5.85E-01 | 4.52E-01 | 3.02E-01 | 4.00E-01 | 6.27E-01 | 7.36E-01 |
| 50 | 3 | 3.20E-01 | 9.96E-02 | 4.66E-01 | 1.27E-02 | 3.98E-02 | 1.14E-01 | 8.43E-02 |
| 50 | 4 | 3.32E-03 | 7.20E-04 | 1.60E-02 | 0.00E+00 | 6.00E-05 | 0.00E+00 | 0.00E+00 |
| 100 | 0 | 1.13E-02 | 1.34E-02 | 2.85E-03 | 1.54E-02 | 9.60E-03 | 9.13E-03 | 6.71E-04 |
| 100 | 1 | 2.66E-02 | 3.60E-02 | 8.57E-03 | 6.95E-02 | 3.55E-02 | 2.76E-02 | 3.26E-03 |
| 100 | 2 | 5.09E-02 | 8.10E-02 | 2.00E-02 | 2.64E-01 | 1.79E-01 | 7.33E-02 | 1.53E-02 |
| 100 | 3 | 1.15E-01 | 2.59E-01 | 6.43E-02 | 4.21E-01 | 4.07E-01 | 2.49E-01 | 1.49E-01 |
| 100 | 4 | 2.93E-01 | 4.05E-01 | 2.32E-01 | 2.04E-01 | 2.96E-01 | 4.19E-01 | 5.54E-01 |
| 100 | 5 | 3.69E-01 | 1.84E-01 | 4.04E-01 | 2.54E-02 | 6.88E-02 | 2.05E-01 | 2.70E-01 |
| 100 | 6 | 1.28E-01 | 2.23E-02 | 2.39E-01 | 7.00E-04 | 4.17E-03 | 1.65E-02 | 7.56E-03 |
| 100 | 7 | 6.65E-03 | 6.20E-04 | 2.97E-02 | 0.00E+00 | 1.00E-05 | 0.00E+00 | 4.94E-06 |
| 300 | 0 | 7.70E-02 | 8.73E-02 | 3.77E-02 | 6.66E-02 | 7.66E-02 | 6.76E-02 | 1.82E-02 |
| 300 | 1 | 1.09E-01 | 1.32E-01 | 7.62E-02 | 1.21E-01 | 1.05E-01 | 1.17E-01 | 5.73E-02 |
| 300 | 2 | 1.20E-01 | 1.39E-01 | 9.99E-02 | 1.39E-01 | 1.12E-01 | 1.30E-01 | 9.85E-02 |
| 300 | 3 | 1.13E-01 | 1.27E-01 | 1.02E-01 | 1.32E-01 | 1.04E-01 | 1.21E-01 | 1.22E-01 |
| 300 | 4 | 9.99E-02 | 1.05E-01 | 9.67E-02 | 1.13E-01 | 9.36E-02 | 1.04E-01 | 1.23E-01 |
| 300 | 5 | 8.38E-02 | 8.46E-02 | 8.49E-02 | 9.48E-02 | 8.06E-02 | 8.53E-02 | 1.10E-01 |
| 300 | 6 | 6.98E-02 | 6.65E-02 | 7.39E-02 | 7.63E-02 | 6.94E-02 | 6.94E-02 | 9.08E-02 |
| 300 | 7 | 5.76E-02 | 5.32E-02 | 6.21E-02 | 6.63E-02 | 6.35E-02 | 5.45E-02 | 7.17E-02 |
| 300 | 8 | 4.77E-02 | 4.26E-02 | 5.27E-02 | 5.95E-02 | 6.40E-02 | 4.71E-02 | 5.59E-02 |
| 300 | 9 | 4.01E-02 | 3.62E-02 | 4.46E-02 | 5.22E-02 | 6.55E-02 | 4.24E-02 | 4.51E-02 |
| 300 | 10 | 3.50E-02 | 3.36E-02 | 3.91E-02 | 4.17E-02 | 6.30E-02 | 4.12E-02 | 3.94E-02 |
| 300 | 11 | 3.15E-02 | 3.23E-02 | 3.63E-02 | 2.49E-02 | 5.30E-02 | 4.16E-02 | 4.13E-02 |
| 300 | 12 | 3.01E-02 | 2.83E-02 | 3.42E-02 | 9.80E-03 | 3.19E-02 | 3.68E-02 | 4.92E-02 |
| 300 | 13 | 2.88E-02 | 1.94E-02 | 3.77E-02 | 2.60E-03 | 1.36E-02 | 2.55E-02 | 4.71E-02 |
| 300 | 14 | 2.48E-02 | 9.64E-03 | 3.88E-02 | 4.00E-04 | 3.68E-03 | 1.21E-02 | 2.44E-02 |
| 300 | 15 | 1.82E-02 | 2.96E-03 | 3.57E-02 | 1.00E-04 | 6.50E-04 | 3.38E-03 | 5.05E-03 |
| 300 | 16 | 9.61E-03 | 6.80E-04 | 2.70E-02 | 0.00E+00 | 7.00E-05 | 6.50E-04 | 3.43E-04 |
| 300 | 17 | 3.12E-03 | 4.00E-05 | 1.42E-02 | 0.00E+00 | 0.00E+00 | 4.00E-05 | 1.71E-05 |
| 300 | 18 | 5.40E-04 | 1.00E-05 | 5.10E-03 | 0.00E+00 | 0.00E+00 | 0.00E+00 | 0.00E+00 |
| 300 | 19 | 1.00E-04 | 0.00E+00 | 1.26E-03 | 0.00E+00 | 0.00E+00 | 0.00E+00 | 0.00E+00 |
| 300 | 20 | 0.00E+00 | 0.00E+00 | 1.70E-04 | 0.00E+00 | 0.00E+00 | 0.00E+00 | 0.00E+00 |
| 600 | 0 | 2.12E-01 | 2.20E-01 | 1.35E-01 | 2.02E-01 | 2.13E-01 | 1.94E-01 | 8.06E-02 |

| | | | | | | | |
|---|---|---|---|---|---|---|---|
| 600 | 1 | 1.98E-01 | 2.25E-01 | 1.81E-01 | 2.33E-01 | 1.95E-01 | 2.23E-01 | 1.70E-01 |
| 600 | 2 | 1.66E-01 | 1.76E-01 | 1.70E-01 | 1.87E-01 | 1.58E-01 | 1.78E-01 | 1.95E-01 |
| 600 | 3 | 1.22E-01 | 1.23E-01 | 1.34E-01 | 1.32E-01 | 1.19E-01 | 1.24E-01 | 1.65E-01 |
| 600 | 4 | 8.64E-02 | 8.28E-02 | 1.01E-01 | 8.27E-02 | 8.61E-02 | 8.56E-02 | 1.17E-01 |
| 600 | 5 | 6.11E-02 | 5.45E-02 | 7.38E-02 | 5.53E-02 | 6.27E-02 | 5.91E-02 | 7.92E-02 |
| 600 | 6 | 4.23E-02 | 3.68E-02 | 5.33E-02 | 3.49E-02 | 4.48E-02 | 3.97E-02 | 5.11E-02 |
| 600 | 7 | 3.03E-02 | 2.51E-02 | 3.76E-02 | 2.28E-02 | 3.14E-02 | 2.68E-02 | 3.35E-02 |
| 600 | 8 | 2.28E-02 | 1.76E-02 | 2.75E-02 | 1.56E-02 | 2.37E-02 | 1.93E-02 | 2.40E-02 |
| 600 | 9 | 1.54E-02 | 1.19E-02 | 2.09E-02 | 1.07E-02 | 1.68E-02 | 1.35E-02 | 1.73E-02 |
| 600 | 10 | 1.08E-02 | 8.55E-03 | 1.57E-02 | 7.60E-03 | 1.17E-02 | 1.02E-02 | 1.29E-02 |
| 600 | 11 | 8.23E-03 | 5.78E-03 | 1.13E-02 | 4.80E-03 | 9.14E-03 | 6.91E-03 | 1.07E-02 |
| 600 | 12 | 6.26E-03 | 3.85E-03 | 8.66E-03 | 3.50E-03 | 6.35E-03 | 5.25E-03 | 8.70E-03 |
| 600 | 13 | 4.51E-03 | 2.85E-03 | 6.79E-03 | 2.20E-03 | 5.10E-03 | 3.90E-03 | 7.41E-03 |
| 600 | 14 | 3.50E-03 | 2.05E-03 | 5.18E-03 | 1.70E-03 | 4.25E-03 | 2.99E-03 | 5.76E-03 |
| 600 | 15 | 2.33E-03 | 1.63E-03 | 4.17E-03 | 1.50E-03 | 3.06E-03 | 2.10E-03 | 4.91E-03 |
| 600 | 16 | 1.92E-03 | 9.30E-04 | 3.19E-03 | 1.10E-03 | 2.49E-03 | 1.53E-03 | 3.78E-03 |
| 600 | 17 | 1.69E-03 | 7.10E-04 | 2.54E-03 | 7.00E-04 | 1.83E-03 | 1.27E-03 | 3.28E-03 |
| 600 | 18 | 1.22E-03 | 5.20E-04 | 1.97E-03 | 5.00E-04 | 1.51E-03 | 9.20E-04 | 2.48E-03 |
| 600 | 19 | 7.90E-04 | 4.20E-04 | 1.66E-03 | 3.00E-04 | 1.22E-03 | 6.20E-04 | 2.05E-03 |
| 600 | 20 | 7.50E-04 | 2.60E-04 | 1.28E-03 | 3.00E-04 | 8.10E-04 | 4.10E-04 | 1.74E-03 |
| 600 | 21 | 5.40E-04 | 1.10E-04 | 1.07E-03 | 1.00E-04 | 5.40E-04 | 5.60E-04 | 1.24E-03 |
| 600 | 22 | 3.40E-04 | 1.50E-04 | 8.40E-04 | 1.00E-04 | 3.90E-04 | 2.40E-04 | 9.46E-04 |
| 600 | 23 | 2.70E-04 | 1.50E-04 | 6.80E-04 | 0.00E+00 | 1.80E-04 | 2.20E-04 | 8.94E-04 |
| 600 | 24 | 2.90E-04 | 9.00E-05 | 5.70E-04 | 0.00E+00 | 1.00E-04 | 1.80E-04 | 7.06E-04 |
| 600 | 25 | 1.50E-04 | 5.00E-05 | 5.70E-04 | 0.00E+00 | 4.00E-05 | 1.40E-04 | 5.63E-04 |
| 600 | 26 | 1.20E-04 | 4.00E-05 | 3.80E-04 | 0.00E+00 | 2.00E-05 | 5.00E-05 | 4.47E-04 |
| 600 | 27 | 1.40E-04 | 2.00E-05 | 2.60E-04 | 0.00E+00 | 0.00E+00 | 1.00E-05 | 2.85E-04 |
| 600 | 28 | 4.00E-05 | 0.00E+00 | 2.60E-04 | 0.00E+00 | 0.00E+00 | 0.00E+00 | 1.30E-04 |

| Energy (keV) | ICS | G4-DNA Opt 2 | G4-DNA Opt 4 | G4-DNA Opt 6 | PARTRAC | MCwater | PTra | PHITS |
|---|---|---|---|---|---|---|---|---|
| 1 | 0 | 3.54E-01 | 3.53E-01 | 2.58E-01 | 3.47E-01 | 3.45E-01 | 3.31E-01 | 1.88E-01 |
| 1 | 1 | 2.29E-01 | 2.54E-01 | 2.40E-01 | 2.72E-01 | 2.27E-01 | 2.60E-01 | 2.74E-01 |
| 1 | 2 | 1.52E-01 | 1.57E-01 | 1.74E-01 | 1.65E-01 | 1.52E-01 | 1.63E-01 | 2.21E-01 |
| 1 | 3 | 9.51E-02 | 9.27E-02 | 1.12E-01 | 9.20E-02 | 9.82E-02 | 9.48E-02 | 1.34E-01 |
| 1 | 4 | 5.93E-02 | 5.47E-02 | 7.23E-02 | 5.09E-02 | 6.34E-02 | 5.55E-02 | 7.53E-02 |
| 1 | 5 | 3.79E-02 | 3.29E-02 | 4.62E-02 | 2.85E-02 | 3.92E-02 | 3.41E-02 | 3.90E-02 |
| 1 | 6 | 2.43E-02 | 2.02E-02 | 3.05E-02 | 1.71E-02 | 2.54E-02 | 2.16E-02 | 2.22E-02 |
| 1 | 7 | 1.64E-02 | 1.26E-02 | 2.08E-02 | 1.08E-02 | 1.66E-02 | 1.32E-02 | 1.37E-02 |
| 1 | 8 | 1.04E-02 | 8.40E-03 | 1.43E-02 | 6.50E-03 | 1.13E-02 | 9.05E-03 | 9.11E-03 |
| 1 | 9 | 6.92E-03 | 5.19E-03 | 9.55E-03 | 3.90E-03 | 6.91E-03 | 6.14E-03 | 6.05E-03 |
| 1 | 10 | 4.60E-03 | 3.77E-03 | 6.72E-03 | 2.40E-03 | 4.77E-03 | 3.86E-03 | 4.54E-03 |
| 1 | 11 | 3.24E-03 | 1.89E-03 | 4.88E-03 | 1.70E-03 | 3.27E-03 | 2.90E-03 | 3.34E-03 |
| 1 | 12 | 2.14E-03 | 1.34E-03 | 3.25E-03 | 1.00E-03 | 1.94E-03 | 1.81E-03 | 2.16E-03 |
| 1 | 13 | 1.45E-03 | 9.10E-04 | 2.38E-03 | 6.00E-04 | 1.31E-03 | 1.11E-03 | 1.71E-03 |
| 1 | 14 | 1.00E-03 | 6.80E-04 | 1.63E-03 | 5.00E-04 | 9.70E-04 | 7.90E-04 | 1.25E-03 |

| | | | | | | | |
|---|---|---|---|---|---|---|---|
| 1 | 15 | 6.30E-04 | 4.90E-04 | 1.35E-03 | 2.00E-04 | 7.80E-04 | 6.20E-04 | 1.17E-03 |
| 1 | 16 | 4.20E-04 | 1.70E-04 | 7.60E-04 | 1.00E-04 | 4.20E-04 | 4.20E-04 | 6.72E-04 |
| 1 | 17 | 3.20E-04 | 1.40E-04 | 6.20E-04 | 1.00E-04 | 3.60E-04 | 2.80E-04 | 6.18E-04 |
| 1 | 18 | 1.50E-04 | 1.30E-04 | 5.30E-04 | 1.00E-04 | 2.00E-04 | 2.30E-04 | 4.37E-04 |
| 1 | 19 | 1.10E-04 | 7.00E-05 | 3.40E-04 | 1.00E-04 | 1.80E-04 | 1.40E-04 | 4.69E-04 |
| 1 | 20 | 1.40E-04 | 8.00E-05 | 2.50E-04 | 0.00E+00 | 1.40E-04 | 7.00E-05 | 1.92E-04 |
| 1 | 21 | 4.00E-05 | 3.00E-05 | 2.30E-04 | 0.00E+00 | 5.00E-05 | 6.00E-05 | 3.09E-04 |
| 1 | 22 | 1.00E-04 | 2.00E-05 | 9.00E-05 | 0.00E+00 | 2.00E-05 | 3.00E-05 | 6.40E-05 |
| 1 | 23 | 0.00E+00 | 3.00E-05 | 3.00E-05 | 0.00E+00 | 1.00E-05 | 4.00E-05 | 1.81E-04 |
| 1 | 24 | 3.00E-05 | 2.00E-05 | 8.00E-05 | 0.00E+00 | 3.00E-05 | 2.00E-05 | 7.46E-05 |
| 1 | 25 | 3.00E-05 | 1.00E-05 | 5.00E-05 | 0.00E+00 | 2.00E-05 | 0.00E+00 | 7.46E-05 |
| 1 | 26 | 3.00E-05 | 0.00E+00 | 4.00E-05 | 0.00E+00 | 2.00E-05 | 1.00E-05 | 5.33E-05 |
| 1 | 27 | 0.00E+00 | 0.00E+00 | 3.00E-05 | 0.00E+00 | 0.00E+00 | 0.00E+00 | 4.26E-05 |
| 1 | 28 | 0.00E+00 | 1.00E-05 | 1.00E-05 | 0.00E+00 | 2.00E-05 | 0.00E+00 | 6.40E-05 |
| 5 | 0 | 3.34E-02 | 2.95E-02 | 1.18E-02 | 2.75E-02 | 3.65E-02 | 2.55E-02 | 4.52E-03 |
| 5 | 1 | 6.33E-02 | 6.89E-02 | 3.41E-02 | 7.07E-02 | 7.29E-02 | 6.63E-02 | 1.87E-02 |
| 5 | 2 | 9.06E-02 | 9.90E-02 | 6.07E-02 | 1.04E-01 | 9.83E-02 | 9.84E-02 | 4.79E-02 |
| 5 | 3 | 1.01E-01 | 1.13E-01 | 8.13E-02 | 1.16E-01 | 1.05E-01 | 1.12E-01 | 8.01E-02 |
| 5 | 4 | 1.00E-01 | 1.11E-01 | 9.41E-02 | 1.09E-01 | 1.04E-01 | 1.11E-01 | 1.03E-01 |
| 5 | 5 | 9.28E-02 | 1.02E-01 | 9.51E-02 | 9.89E-02 | 9.70E-02 | 1.02E-01 | 1.15E-01 |
| 5 | 6 | 8.31E-02 | 8.74E-02 | 9.23E-02 | 8.28E-02 | 8.11E-02 | 8.58E-02 | 1.13E-01 |
| 5 | 7 | 7.00E-02 | 7.23E-02 | 8.34E-02 | 6.57E-02 | 6.89E-02 | 7.10E-02 | 1.02E-01 |
| 5 | 8 | 6.06E-02 | 5.69E-02 | 7.34E-02 | 5.11E-02 | 5.52E-02 | 5.77E-02 | 8.20E-02 |
| 5 | 9 | 5.02E-02 | 4.58E-02 | 6.10E-02 | 4.15E-02 | 4.44E-02 | 4.63E-02 | 6.50E-02 |
| 5 | 10 | 4.07E-02 | 3.77E-02 | 5.16E-02 | 3.19E-02 | 3.66E-02 | 3.53E-02 | 4.59E-02 |
| 5 | 11 | 3.43E-02 | 3.02E-02 | 4.33E-02 | 2.54E-02 | 2.87E-02 | 2.90E-02 | 3.57E-02 |
| 5 | 12 | 2.78E-02 | 2.60E-02 | 3.63E-02 | 2.06E-02 | 2.32E-02 | 2.37E-02 | 2.78E-02 |
| 5 | 13 | 2.36E-02 | 2.01E-02 | 2.90E-02 | 1.60E-02 | 1.85E-02 | 1.89E-02 | 2.02E-02 |
| 5 | 14 | 1.97E-02 | 1.61E-02 | 2.42E-02 | 1.39E-02 | 1.55E-02 | 1.50E-02 | 1.51E-02 |
| 5 | 15 | 1.59E-02 | 1.33E-02 | 1.99E-02 | 1.11E-02 | 1.28E-02 | 1.29E-02 | 1.09E-02 |
| 5 | 16 | 1.40E-02 | 1.08E-02 | 1.78E-02 | 9.70E-03 | 1.05E-02 | 1.05E-02 | 9.37E-03 |
| 5 | 17 | 1.27E-02 | 9.24E-03 | 1.49E-02 | 8.20E-03 | 8.83E-03 | 9.09E-03 | 6.42E-03 |
| 5 | 18 | 1.01E-02 | 8.16E-03 | 1.24E-02 | 7.30E-03 | 8.09E-03 | 7.66E-03 | 5.30E-03 |
| 5 | 19 | 8.90E-03 | 6.83E-03 | 1.05E-02 | 6.60E-03 | 6.70E-03 | 6.33E-03 | 5.22E-03 |
| 5 | 20 | 7.81E-03 | 5.70E-03 | 8.23E-03 | 5.70E-03 | 6.05E-03 | 5.71E-03 | 4.81E-03 |
| 5 | 21 | 6.77E-03 | 4.99E-03 | 8.21E-03 | 5.20E-03 | 5.06E-03 | 5.16E-03 | 3.47E-03 |
| 5 | 22 | 5.63E-03 | 4.65E-03 | 6.62E-03 | 5.30E-03 | 4.28E-03 | 3.89E-03 | 3.15E-03 |
| 5 | 23 | 5.29E-03 | 3.91E-03 | 5.70E-03 | 4.70E-03 | 3.77E-03 | 3.99E-03 | 2.92E-03 |
| 5 | 24 | 4.65E-03 | 3.87E-03 | 5.41E-03 | 4.30E-03 | 3.27E-03 | 3.33E-03 | 2.87E-03 |
| 5 | 25 | 4.04E-03 | 3.26E-03 | 4.48E-03 | 3.80E-03 | 3.32E-03 | 3.18E-03 | 2.06E-03 |
| 5 | 26 | 3.70E-03 | 2.65E-03 | 3.98E-03 | 4.00E-03 | 2.93E-03 | 2.60E-03 | 2.32E-03 |
| 5 | 27 | 3.45E-03 | 2.69E-03 | 3.78E-03 | 3.60E-03 | 2.64E-03 | 2.28E-03 | 2.38E-03 |
| 5 | 28 | 3.07E-03 | 2.31E-03 | 3.51E-03 | 3.10E-03 | 2.47E-03 | 2.07E-03 | 2.09E-03 |
| 10 | 0 | 1.50E-01 | 1.19E-01 | 8.35E-02 | 1.41E-01 | 1.53E-01 | 1.36E-01 | 4.84E-02 |
| 10 | 1 | 1.62E-01 | 1.71E-01 | 1.42E-01 | 1.93E-01 | 1.75E-01 | 1.89E-01 | 1.27E-01 |
| 10 | 2 | 1.55E-01 | 1.67E-01 | 1.55E-01 | 1.76E-01 | 1.61E-01 | 1.74E-01 | 1.74E-01 |
| 10 | 3 | 1.27E-01 | 1.36E-01 | 1.39E-01 | 1.33E-01 | 1.28E-01 | 1.35E-01 | 1.77E-01 |
| 10 | 4 | 9.57E-02 | 1.04E-01 | 1.12E-01 | 9.62E-02 | 9.58E-02 | 9.68E-02 | 1.40E-01 |
| 10 | 5 | 7.11E-02 | 7.50E-02 | 8.65E-02 | 6.39E-02 | 6.89E-02 | 6.84E-02 | 9.89E-02 |
| 10 | 6 | 5.23E-02 | 5.31E-02 | 6.48E-02 | 4.28E-02 | 4.81E-02 | 4.69E-02 | 6.46E-02 |
| 10 | 7 | 3.94E-02 | 3.86E-02 | 4.82E-02 | 2.94E-02 | 3.57E-02 | 3.33E-02 | 4.04E-02 |
| 10 | 8 | 2.91E-02 | 2.89E-02 | 3.53E-02 | 2.02E-02 | 2.52E-02 | 2.46E-02 | 2.61E-02 |
| 10 | 9 | 2.27E-02 | 2.13E-02 | 2.69E-02 | 1.48E-02 | 1.92E-02 | 1.76E-02 | 1.65E-02 |
| 10 | 10 | 1.63E-02 | 1.60E-02 | 2.06E-02 | 1.08E-02 | 1.34E-02 | 1.28E-02 | 1.20E-02 |
| 10 | 11 | 1.39E-02 | 1.20E-02 | 1.55E-02 | 8.50E-03 | 1.06E-02 | 1.01E-02 | 8.40E-03 |
| 10 | 12 | 1.09E-02 | 9.46E-03 | 1.25E-02 | 7.30E-03 | 8.36E-03 | 8.06E-03 | 6.11E-03 |
| 10 | 13 | 8.47E-03 | 7.62E-03 | 9.53E-03 | 5.40E-03 | 6.43E-03 | 6.01E-03 | 4.12E-03 |
| 10 | 14 | 6.61E-03 | 6.44E-03 | 7.63E-03 | 5.10E-03 | 5.28E-03 | 4.71E-03 | 4.00E-03 |
| 10 | 15 | 6.24E-03 | 4.98E-03 | 6.26E-03 | 4.00E-03 | 4.88E-03 | 4.18E-03 | 3.22E-03 |
| 10 | 16 | 5.07E-03 | 4.01E-03 | 5.11E-03 | 3.50E-03 | 3.85E-03 | 3.66E-03 | 2.85E-03 |
| 10 | 17 | 4.32E-03 | 3.61E-03 | 4.40E-03 | 2.80E-03 | 3.24E-03 | 2.89E-03 | 2.06E-03 |
| 10 | 18 | 3.64E-03 | 3.30E-03 | 4.14E-03 | 2.90E-03 | 2.70E-03 | 2.71E-03 | 1.78E-03 |
| 10 | 19 | 3.17E-03 | 3.07E-03 | 2.93E-03 | 2.70E-03 | 2.63E-03 | 2.42E-03 | 1.46E-03 |
| 10 | 20 | 2.55E-03 | 2.38E-03 | 3.06E-03 | 2.70E-03 | 2.22E-03 | 2.02E-03 | 1.18E-03 |
| 10 | 21 | 2.57E-03 | 2.26E-03 | 2.65E-03 | 2.50E-03 | 2.18E-03 | 1.72E-03 | 1.02E-03 |
| 10 | 22 | 1.91E-03 | 1.63E-03 | 2.15E-03 | 2.40E-03 | 1.69E-03 | 1.43E-03 | 1.13E-03 |
| 10 | 23 | 1.84E-03 | 1.62E-03 | 1.82E-03 | 2.20E-03 | 1.59E-03 | 1.50E-03 | 1.16E-03 |
| 10 | 24 | 1.91E-03 | 1.60E-03 | 1.82E-03 | 2.10E-03 | 1.49E-03 | 1.22E-03 | 9.95E-04 |
| 10 | 25 | 1.65E-03 | 1.54E-03 | 1.62E-03 | 1.90E-03 | 1.25E-03 | 1.09E-03 | 1.48E-03 |
| 10 | 26 | 1.31E-03 | 1.29E-03 | 1.37E-03 | 1.80E-03 | 1.32E-03 | 1.01E-03 | 1.13E-03 |
| 10 | 27 | 1.35E-03 | 1.19E-03 | 1.07E-03 | 1.50E-03 | 1.46E-03 | 9.60E-04 | 1.25E-03 |
| 10 | 28 | 1.15E-03 | 1.18E-03 | 1.15E-03 | 1.50E-03 | 1.16E-03 | 8.10E-04 | 1.74E-03 |

## Wasserstein distances between ICSDs with original and modified codes

*Supplementary Table S- 2 : Wasserstein distances calculated between each original ICSD and the mean ICSD presented in black color in Figure 1 and Figure 2*

| Energy (eV) | G4-DNA Opt 2 | G4-DNA Opt 4 | G4-DNA Opt 6 | PARTRAC | MCwater | PTra | PHITS | Mean value |
|---|---|---|---|---|---|---|---|---|
| **20** | 0.25 | 0.08 | 0.32 | 0.31 | 0.22 | 0.11 | 0.06 | 0.19 |
| **50** | 0.38 | 0.09 | 0.60 | 0.61 | 0.41 | 0.12 | 0.24 | 0.35 |
| **100** | 0.51 | 0.16 | 1.03 | 0.99 | 0.63 | 0.16 | 0.43 | 0.56 |
| **300** | 0.23 | 0.78 | 1.33 | 0.88 | 0.33 | 0.31 | 0.78 | 0.66 |
| **600** | 0.10 | 0.47 | 0.59 | 0.51 | 0.13 | 0.24 | 0.75 | 0.40 |
| **1000** | 0.09 | 0.21 | 0.38 | 0.31 | 0.09 | 0.12 | 0.32 | 0.22 |
| **5000** | 0.34 | 0.85 | 1.03 | 0.65 | 0.42 | 0.43 | 1.15 | 0.70 |
| **10000** | 0.26 | 0.19 | 0.60 | 0.47 | 0.24 | 0.37 | 0.65 | 0.40 |

Supplementary Table S- 3 : Wasserstein distances calculated between each ICSD obtained with modified MCTS codes and the mean ICSD presented in black color in

*Figure 3 and Figure 4*

| Energy (eV) | G4-DNA Opt 2 | G4-DNA Opt 4 | G4-DNA Opt 6 | PARTRAC | MCwater | PTra | PHITS | Mean value |
|---|---|---|---|---|---|---|---|---|
| 20 | 0.013 | 0.002 | 0.002 | 0.01 | 0.007 | 0.002 | 0.012 | 0.007 |
| 50 | 0.04 | 0.02 | 0.10 | 0.04 | 0.15 | 0.09 | 0.13 | 0.08 |
| 100 | 0.09 | 0.04 | 0.17 | 0.12 | 0.34 | 0.24 | 0.29 | 0.18 |
| 300 | 0.50 | 0.16 | 0.33 | 0.13 | 0.31 | 0.66 | 0.16 | 0.32 |
| 600 | 0.26 | 0.22 | 0.39 | 0.26 | 0.32 | 0.62 | 0.53 | 0.37 |
| 1000 | 0.25 | 0.24 | 0.23 | 0.26 | 0.29 | 0.48 | 0.11 | 0.27 |
| 5000 | 0.25 | 0.30 | 0.77 | 0.44 | 0.60 | 1.34 | 0.63 | 0.62 |
| 10000 | 0.27 | 0.30 | 0.33 | 0.31 | 0.48 | 0.96 | 0.16 | 0.40 |

*Supplementary Table S- 4 : Wasserstein distances calculated between each ICSD obtained with the original MCTS code and its modified version using common cross sections used in this work*

| Wasserstein-1 distances, original- modified MCTS codes | | | | | | | |
|---|---|---|---|---|---|---|---|
| Energy (eV) | G4-DNA Opt 2 | G4-DNA Opt 4 | G4-DNA Opt 6 | PARTRAC | MCwater | PTra | PHITS |
| 20 | 0.17 | 0.17 | 0.22 | 0.41 | 0.31 | 0.01 | 0.47 |
| 50 | 0.38 | 0.08 | 0.49 | 0.64 | 0.34 | 0.08 | 1.20 |
| 100 | 0.58 | 0.08 | 0.95 | 1.02 | 0.42 | 0.25 | 2.29 |
| 300 | 0.61 | 0.71 | 1.42 | 1.66 | 0.25 | 0.44 | 2.79 |
| 600 | 0.47 | 0.69 | 0.80 | 0.41 | 0.39 | 0.27 | 2.22 |
| 1000 | 0.48 | 0.62 | 0.38 | 0.26 | 0.42 | 0.19 | 1.30 |
| 5000 | 0.92 | 1.52 | 1.46 | 0.75 | 1.19 | 0.54 | 4.88 |
| 10000 | 0.77 | 0.85 | 0.63 | 0.51 | 0.94 | 0.23 | 2.75 |

# Nanodosimetric quantities results: $F_3$ values (cumulative probability of obtaining 3 or more ionizations)

*Supplementary Table S- 5 : $F_3$ values obtained with original MCTS codes, and standard deviation (SD) and relative standard deviation (RSD) of this set of codes*

| $F_3$: Probability of three or more ionizations predicted by the original MCTS codes | | | | | | | | | |
|---|---|---|---|---|---|---|---|---|---|
| Energy (eV) | G4-DNA Opt 2 | G4-DNA Opt 4 | G4-DNA Opt 6 | PARTRAC | MCwater | PTra | PHITS | SD | RSD |
| **50** | 0.32 | 0.10 | 0.48 | 0.01 | 0.04 | 0.11 | 0.08 | 0.17 | 1.04 |
| **100** | 0.91 | 0.87 | 0.97 | 0.65 | 0.78 | 0.89 | 0.98 | 0.12 | 0.13 |
| **300** | 0.69 | 0.64 | 0.79 | 0.67 | 0.71 | 0.69 | 0.83 | 0.07 | 0.09 |
| **600** | 0.42 | 0.38 | 0.52 | 0.38 | 0.43 | 0.41 | 0.56 | 0.07 | 0.15 |
| **1000** | 0.26 | 0.24 | 0.33 | 0.22 | 0.28 | 0.25 | 0.32 | 0.04 | 0.15 |
| **5000** | 0.81 | 0.80 | 0.89 | 0.76 | 0.76 | 0.79 | 0.87 | 0.05 | 0.06 |
| **10000** | 0.53 | 0.46 | 0.62 | 0.53 | 0.50 | 0.51 | 0.62 | 0.06 | 0.11 |

*Supplementary Table S- 6 : $F_3$ values obtained with modified MCTS codes using the common cross section and standard deviation (SD) and relative standard deviation (RSD) of this set of codes.*

| $F_3$: Probability of three or more ionizations predicted by the modified MCTS codes | | | | | | | | | |
|---|---|---|---|---|---|---|---|---|---|
| Energy (eV) | G4-DNA Opt 2 | G4-DNA Opt 4 | G4-DNA Opt 6 | PARTRAC | MCwater | PTra | PHITS | SD | RSD |
| **50** | 0.14 | 0.16 | 0.22 | 0.16 | 0.17 | 0.10 | 0.22 | 0.04 | 0.25 |
| **100** | 0.89 | 0.86 | 0.88 | 0.90 | 0.74 | 0.83 | 0.91 | 0.06 | 0.07 |
| **300** | 0.76 | 0.73 | 0.68 | 0.70 | 0.73 | 0.65 | 0.71 | 0.03 | 0.05 |
| **600** | 0.50 | 0.49 | 0.43 | 0.42 | 0.49 | 0.37 | 0.49 | 0.05 | 0.11 |
| **1000** | 0.33 | 0.34 | 0.28 | 0.26 | 0.35 | 0.22 | 0.29 | 0.05 | 0.16 |
| **5000** | 0.86 | 0.86 | 0.83 | 0.82 | 0.85 | 0.78 | 0.87 | 0.03 | 0.04 |
| **10000** | 0.61 | 0.61 | 0.57 | 0.54 | 0.62 | 0.48 | 0.59 | 0.05 | 0.09 |

*Supplementary Table S- 7 : Mean value of the ICSDs with modified options in Geant4-DNA using the common interaction cross sections data set and the same differential elastic cross sections (those originally implemented in option 2, Champion's model). Therefore, the only difference between all those "options" are the differential ionization cross sections and angular distributions in the ionization process.*

| Energy (eV) | G4-DNA opt2 with common cross sections | G4-DNA opt4 with common cross sections | G4-DNA opt4 with common cross sections and differential elastics of opt2 | G4-DNA opt6 with common cross sections | G4-DNA opt6 with common cross sections and differential elastics of opt2 |
|---|---|---|---|---|---|
| 20 | 0.60 | 0.61 | 0.61 | 0.61 | 0.61 |
| 50 | 1.82 | 1.84 | 1.86 | 1.94 | 1.95 |
| 100 | 3.75 | 3.70 | 3.80 | 3.88 | 3.96 |
| 300 | 5.71 | 5.26 | 5.81 | 5.24 | 5.25 |
| 600 | 3.37 | 3.23 | 3.39 | 2.84 | 2.97 |
| 1000 | 2.31 | 2.29 | 2.32 | 2.04 | 2.07 |
| 5000 | 6.91 | 7.00 | 6.93 | 6.16 | 6.19 |
| 10000 | 4.26 | 4,26 | 4.24 | 3.84 | 3.87 |

q

# Supplementary Figures

# Original codes cross sections and common set cross section data

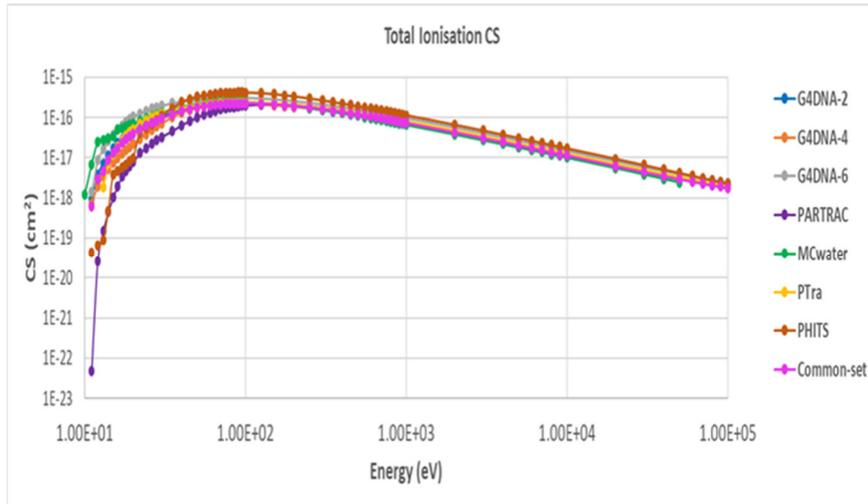

*Supplementary Figure S-1 : Total ionization cross sections originally used in the participating MCTS codes. The data in pink is the total ionization cross section used in the common-set cross sections circulated to the participants for the modification of the codes. The common-set cross sections were calculated as indicated in the materials and methods section.*

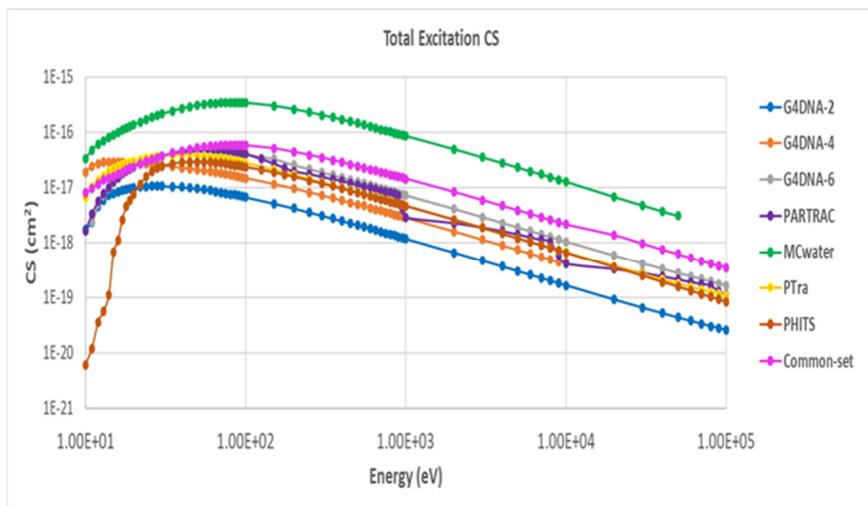

*Supplementary Figure S-2 : Total excitation cross sections of the MCTS codes used by the participants. In pink, the total excitation cross section is shown that was used in the common-set cross sections circulated to the participants for the modification of the codes. The common set cross sections were calculated as indicated in the materials and methods section.*

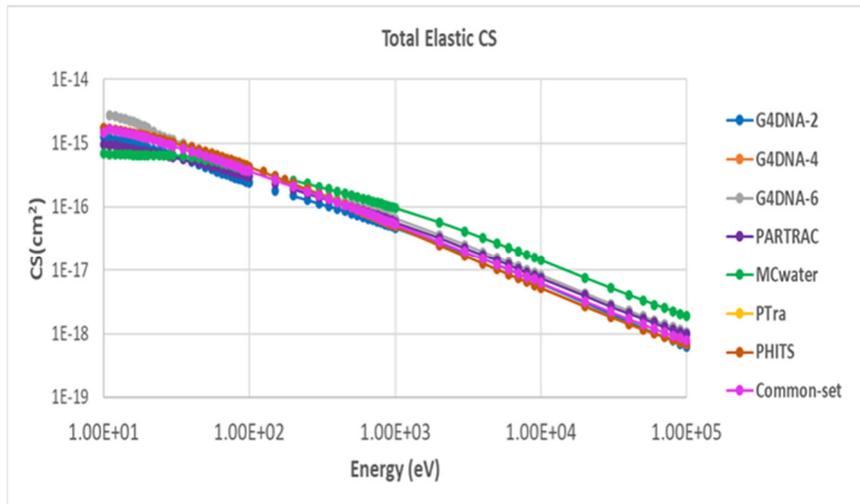

*Supplementary Figure S-3 : Total elastic cross sections of the MCTS codes used by the participants. In pink, the total elastic cross section is shown that was used in the common-set cross sections circulated to the participants for the modification of the codes. The common set cross sections were calculated as indicated in the materials and methods section.*

# Modification of Geant4-DNA options: common cross section data set and equal differential elastic cross sections

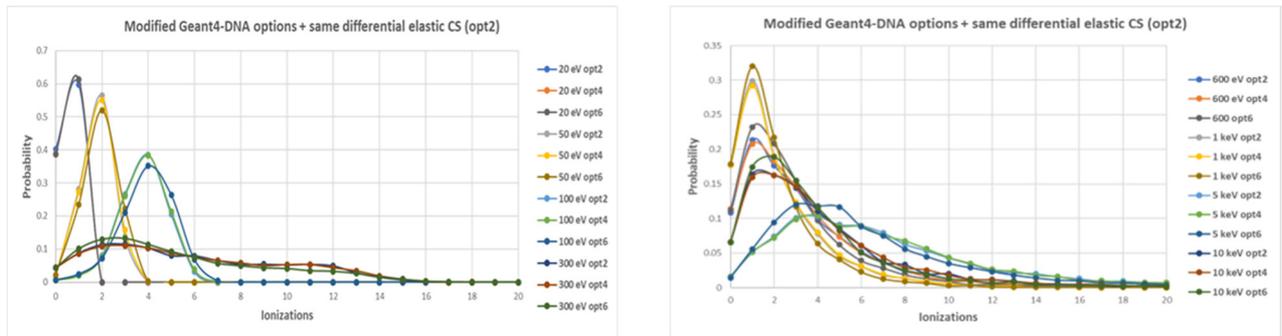

*Supplementary Figure S-4: ICSDs obtained using Geant4-DNA with the common cross section data set, same differential elastic cross sections (those implemented in opt 2 in the original code, Model from Champion) and different ionization cross sections (originally in options 2, 4 and 6)*